\documentclass[aps,prl,reprint,superscriptaddress]{revtex4-1}

\usepackage{amsmath} 
\usepackage{array} 
\usepackage{accents}
\usepackage[super]{nth}
\usepackage{graphicx}
\usepackage{float}
\usepackage[version=4]{mhchem}
\usepackage{siunitx}
\usepackage{placeins}

\newcommand{\ind}[1]{\mathrm{#1''\!\!}}
\newcommand{\direct}[1]{\mathrm{#1'\!}}

\begin{document}

\title{
Indirect Learning of Interatomic Potentials for Accelerated Materials Simulations
}

\author{Joe D. Morrow}
\author{Volker L. Deringer}
\email{volker.deringer@chem.ox.ac.uk}
\affiliation{Department of Chemistry, Inorganic Chemistry Laboratory, University of Oxford, Oxford OX1 3QR, United Kingdom}

\begin{abstract}
    Machine learning (ML) based interatomic potentials are emerging tools for materials simulations but require a trade-off between accuracy and speed. 
    Here we show how one can use one ML potential model to train another: 
    we use an existing, accurate, but more computationally expensive model to generate reference data (locations and labels) for a series of much faster potentials. 
    Without the need for quantum-mechanical reference computations at the secondary stage, extensive reference datasets can be easily generated, and we find that this improves the quality of fast potentials with less flexible functional forms.
    We apply the technique to disordered silicon, including a simulation of vitrification and polycrystalline grain formation under pressure with a system size of a million atoms.
    Our work provides conceptual insight into the machine learning of interatomic potential models, and it suggests a route toward accelerated simulations of condensed-phase systems and nanostructured materials.
\end{abstract}

\maketitle

The properties of materials are governed by the atomic structure and by the forces acting on atoms.
Creating accurate computational models for interatomic forces and potential-energy surfaces is therefore a central task in the physical sciences.
Today, machine learning (ML) methods are increasingly used to represent quantum-mechanical potential-energy surfaces, typically based on density-functional theory (DFT) ground-truth data and achieving prediction accuracy to within a few \si{meV} per atom \cite{Zuo2020}. This way, quantum-mechanical quality simulations become accessible that are orders of magnitude cheaper than DFT \cite{Behler2007, Bartok2010, Thompson2015, Li2015, Shapeev2016, Zhang2018, Jinnouchi2019}.  ML potential models have begun to be applied to challenging problems in physics and related fields: structures and properties of amorphous solids \cite{Sosso2013, Caro2018, aSi_100k}, phase transformations under extreme conditions \cite{Cheng2020, Smith2021, Zong2021}, or surface science and catalysis \cite{Artrith2014, Timmermann2020}. Recent overviews summarize progress in the field \cite{Behler2017, MLP_AdvMater, Noe2020, Friederich2021}.

However, ML potentials still face challenges. A key one is that they incur much larger computational cost than efficient empirical counterparts. 
In addition to developing accurate and fast potentials, there is a similarly important need for {\em validating} ML potential models. This is challenging because there is less physical information, or even none, encoded into the model {\em a priori}, and because there is not always an unambiguous correlation between the numerical error of a potential (on a fixed test set) and its behavior in simulations \cite{George2020}. The importance of physically-motivated tests has been demonstrated, e.g., in a comparative study of carbon interatomic potentials \cite{deTomas2019}. A recent benchmark study compared different ML potential fitting frameworks and their numerical performance \cite{Zuo2020}. Specifically for silicon (Si), various tests including random search have been devised \cite{Bartok2018, George2020, Lysogorskiy2021}.

In the present work, we outline an approach to training ML potentials by {\em indirect learning}: we use one model to create reference data for another. The former is accurate and flexible, but more computationally expensive than the latter. We demonstrate that training a potential in this way for disordered Si largely retains the quality of prediction of the initial model (both in terms of numerical errors and physically motivated tests), whilst speeding up the simulation substantially. This way, much larger-scale simulations become accessible---or conversely, the cost of a given simulation is reduced to a few percent. 

We briefly discuss terminology. The approach herein is related to ``knowledge distillation'' in ML research \cite{Hinton2015}, which seeks to transfer information from a large ``teacher'' model to a smaller ``student'' model, enabling faster predictions. So far, distillation has been applied primarily to classification tasks performed by neural networks, such as image recognition \cite{Hinton2015}. ``Database distillation'', synonymously with ``dataset meta-learning'', was recently described as a way to condense information into smaller training sets for cheaper learning \cite{Nguyen2021}, and the term ``meta-labeling'' has been used to describe the construction of secondary ML models using a primary one \cite{LopezdePrado2018}.  We here think of ``distillation'' in a rather broad sense: to encapsulate the indirect learning of potential models via any fitting framework, and the generation of entirely new atomic configurations for training an indirectly-learned model, in addition to the energy and force data. In other words, we use the teacher model to generate both the data locations and the data labels. 

\begin{figure}[t]
\includegraphics[width=7.75cm]{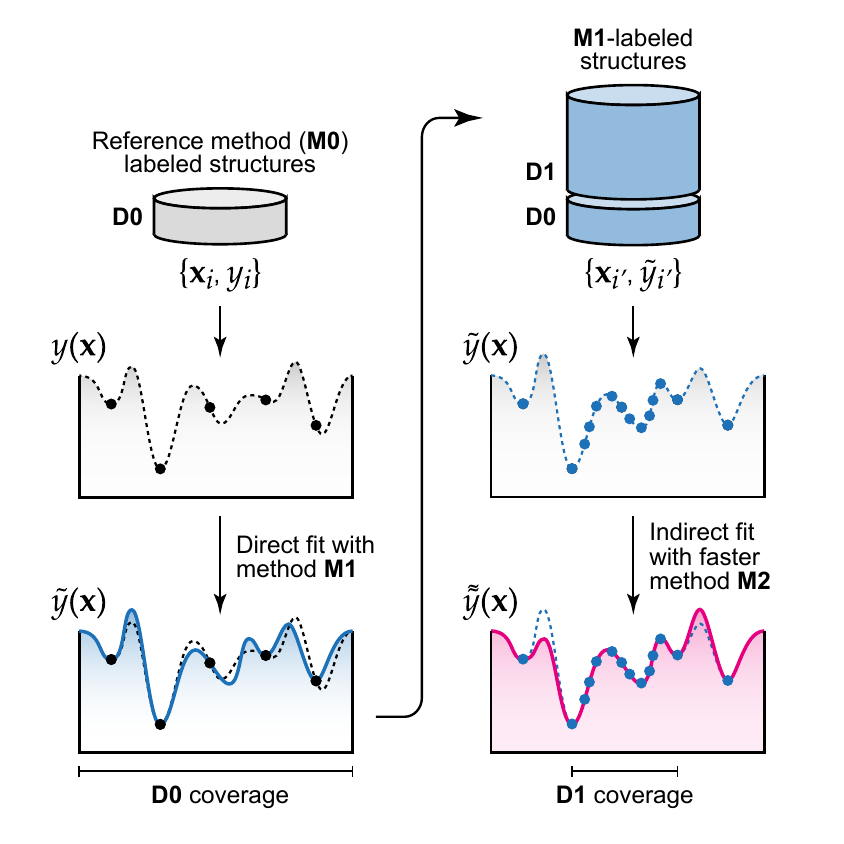}
\caption{
Indirect learning of interatomic potentials. 
\textit{Left:} 
In established ML potentials, a potential-energy surface is represented by reference computations for a given number of structures, leading to the dataset \textbf{D0}. Fitting with the method \textbf{M1} yields an ML potential model (blue line).
\textit{Right:}
We now generate a much larger set of structures using \textbf{M1}, and label them using the same method (not \textbf{M0}!). This dataset, \textbf{D1}, covers a more narrow structural space, but at more points. A second fit is then made with a different ML method, \textbf{M2}.
}
\label{fig:schematic}
\end{figure}

\textit{Methodology.}---ML potentials approximate a potential-energy surface (PES) based on reference data, typically from small-scale DFT computations \cite{chemrev}:
\begin{equation}\label{eq:tilde}
    \left\{ {\bf x}_{i}, y_{i} \right\}_{i=1..N} \xrightarrow{\quad \textbf{M1} \quad} \tilde{y}({\bf x}), 
\end{equation}
where ${\bf x}_{i}$ is a descriptor for the atomic environment \cite{Musil2021}, $y_{i}$ is one of $N$ observations of the reference PES, and $\tilde{y}$ denotes the ML model fitted with a given method \footnote{In practice, the matter is more complex: typically, atomic energies are machine-learned from total energies (and, often, forces) \cite{Behler2007, Bartok2010}.}. We call this method ``\textbf{M1}'' in the following.

\begin{figure}[t]
\includegraphics[width=7.75cm]{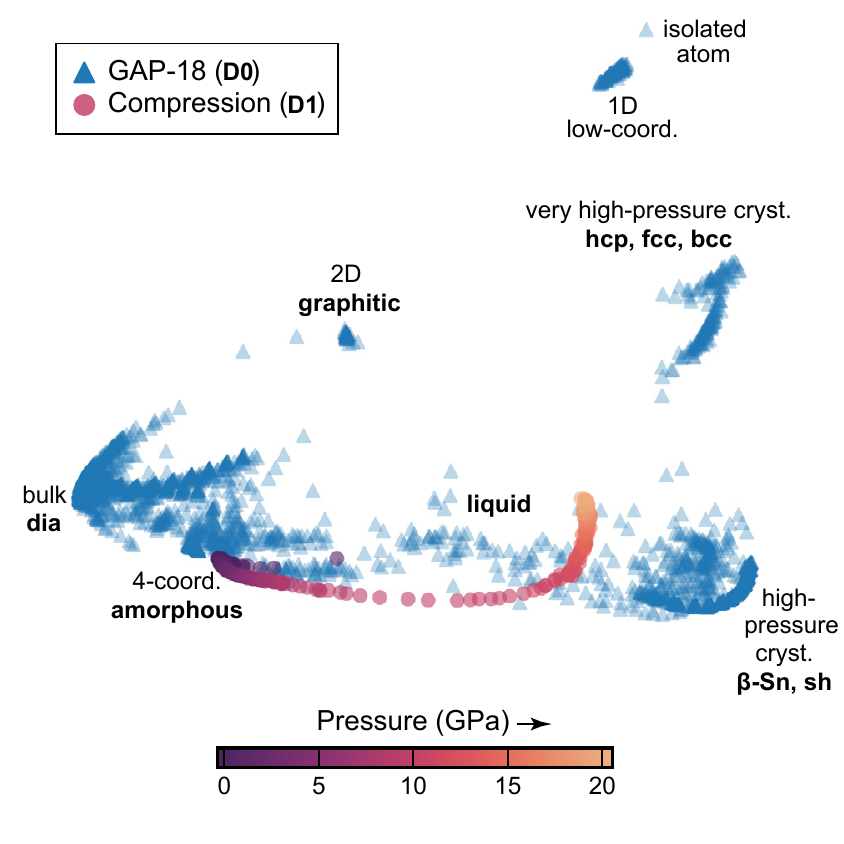}
\caption{
Composition of the ML potential fitting databases used in this work. A SOAP/KPCA map (cf.\ Ref.\ \citenum{Cheng2020a}) compares the structural space covered by the GAP-18 database \cite{GAP-18}, here denoted \textbf{D0}, and the newly created indirect learning dataset, \textbf{D1}. The distance between points on this map is a 2D-embedding of their structural dissimilarity as measured by SOAP \cite{Bartok2013}.
At low pressure, a random starting structure, which rapidly transforms to approximate low-density amorphous (LDA) Si, is met with the good coverage provided by liquid and LDA structures in \textbf{D0}. Equally, the high-pressure region is densely sampled with distorted $\beta$-Sn-type and simple hexagonal (sh) structures. However, the \textbf{D0} coverage is distinctly sparse at intermediate pressures around \SI{10}{GPa}.
}
\label{fig:kpca}
\end{figure}

We now show that two fitting methodologies at different points on the ``Pareto front'' of accuracy and cost \cite{Zuo2020}, namely the Gaussian Approximation Potential (GAP) \cite{Bartok2010, chemrev} and Moment Tensor Potential (MTP) \cite{Shapeev2016, Novikov2021} frameworks, can be combined to accelerate simulations of disordered Si. To do so, we train MTPs using an existing general-purpose ML potential for Si \cite{GAP-18}, referred to as GAP-18, as the reference method. We use the MTP formalism (henceforth denoted \textbf{M2}) to approximate the GAP-18 PES (\textbf{M1}), which is itself approximating the DFT PES (\textbf{M0}). Hence, \textbf{M2} is {\em indirectly} learning quantum-mechanical reference data with \textbf{M1} as its teacher (Fig.\ \ref{fig:schematic}). 
Extending the notation of Eq.\ \ref{eq:tilde},
\begin{equation}\label{eq:tildetilde}
    \left\{ {\bf x}_{i}, \tilde{y}_{i} \right\}_{i=1..N_{2}} \xrightarrow{\quad \textbf{M2} \quad} \tilde{\tilde{y}}({\bf x}), 
\end{equation}
where ${\tilde{\tilde{y}}}$ is an indirectly-learned model, and $N_{2} \gg N$. 
In the following, we use the notation $\mathrm{M'}$ for a directly-learned MTP, and $\mathrm{M''}$ for an indirectly-learned one 
\footnote{We note that the idea of mapping a kernel-based ML potential to a fast approximator has been previously explored: via cubic spline interpolation across a pre-calculated grid of points \cite{Glielmo2018, Vandermause2020}.
}.

\begin{figure*}[t]
\includegraphics[width=15cm]{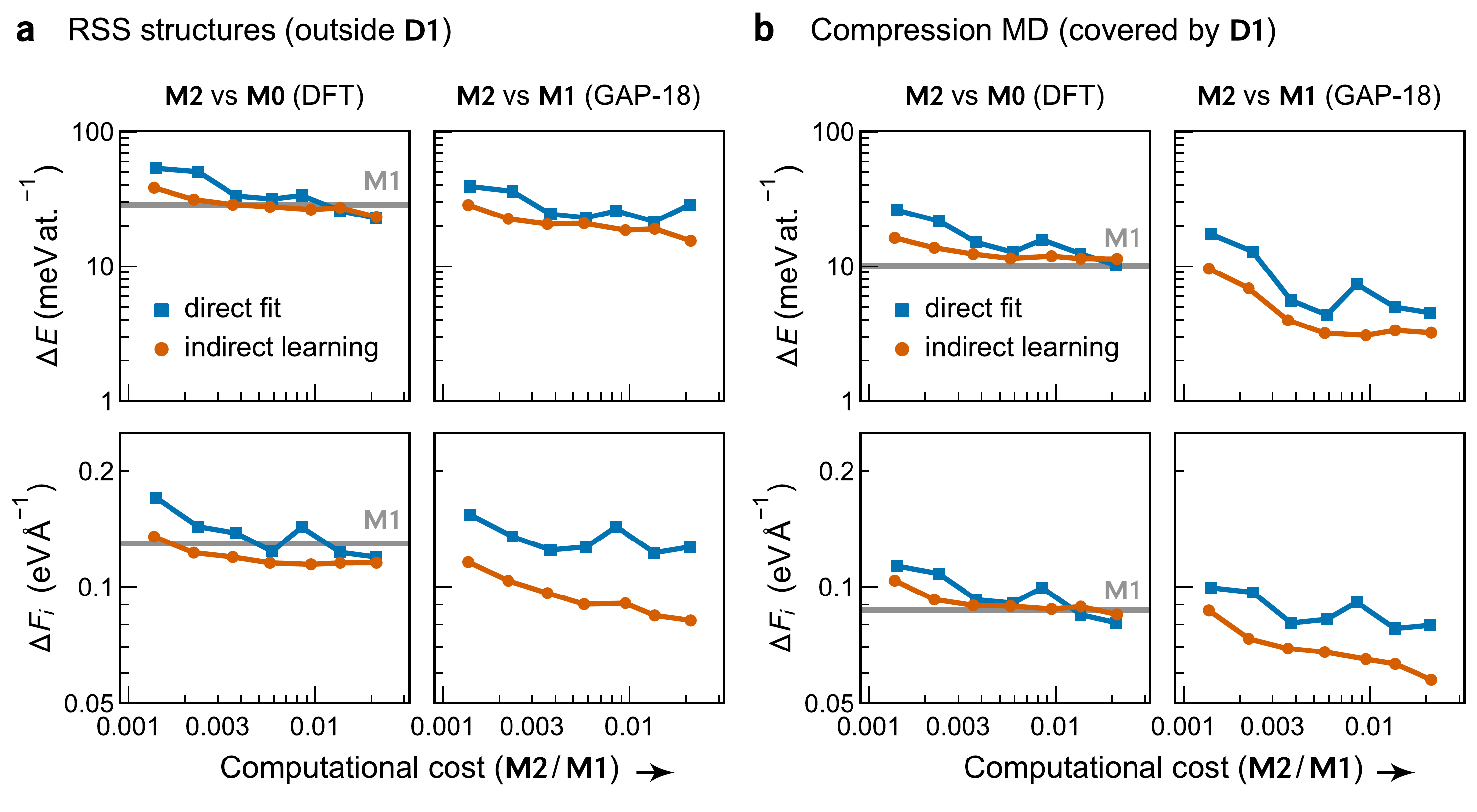}
\caption{Numerical errors versus computational cost for directly- (blue) and indirectly-learned (red) potentials.
The figure characterizes MTPs with increasing number of fitting parameters, expressed via their respective compute-time per MD step relative to GAP-18 (\textbf{M1}). 
(a) Mean absolute error for random structures which were not part of the training. The test comprises 100 cells of 1--64 atoms produced using {\tt buildcell} \cite{Pickard2006, Pickard2011} and optimized with GAP-18 at pressures in the range \SIrange{0}{32}{GPa}. For comparison to \textbf{M0} (\textbf{M1}), test-set structures were labeled with DFT \cite{Perdew1992, Clark2005} (GAP-18 \cite{Bartok2018}), respectively; grey lines report the prediction errors of GAP-18 versus DFT for those structures. This test is not intended as a competitive benchmark, but rather to convey the trade-off between flexibility and efficiency in choosing the number of parameters in MTPs via the maximum level, $L$ (here, set to values of 12--24 in increments of 2). See Supplemental Material for details.
(b) As before, but now for compression MD snapshots: in this case, the configurations {\em have} been included in the indirect learning database, \textbf{D1}.
}\label{fig:rmse}
\end{figure*}

There are two primary advantages to using GAP-18 as an intermediary rather than training directly on DFT-labeled data. Both of these hinge on the massively reduced computational cost of GAP compared to DFT. Firstly, we can build an almost arbitrarily large database of GAP-labeled training data, thus approximating the \textbf{M1} PES as faithfully as possible within the (flexibility and fitting) limitations of \textbf{M2}. Secondly, we can evaluate the fidelity of the fit via numerical errors on a wide-ranging set of out-of-sample testing data, and also test behavior in large-scale molecular dynamics (MD) simulations. The practical advantages of the present application of distillation rely on the observation that GAPs may be able to more accurately interpolate across a sparsely sampled PES compared to MTPs, which is not yet fully investigated \cite{Rosenbrock2021,Zuo2020}. We expect, however, that the same concept could be applied to any pair of fitting methodologies where a cost--accuracy disparity exists. The insights into database development and physically-guided validation of ML potentials are independent of the particular \textbf{M1} and \textbf{M2} methods chosen.

For the present study, we aim to replicate the behavior of GAP-18 under pressure, which we find to be an instructive test case, and which relates to questions of experimental interest \cite{Deb2001, Wilding2006, McMillan2021}. To improve interpolation for high-$p$ structures, we use a large training database (\textbf{D1} in Fig.\ \ref{fig:schematic}) containing 250 structures with 1,000 atoms each, corresponding to compression to 20 GPa (as in Ref.\ \citenum{aSi_100k}). These structures are supplemented with the original GAP-18 database (Fig.\ \ref{fig:kpca}), which we re-label with GAP-18 energies and forces to make the input data consistent. 

\begin{figure*}[t]
    \includegraphics[width=15cm]{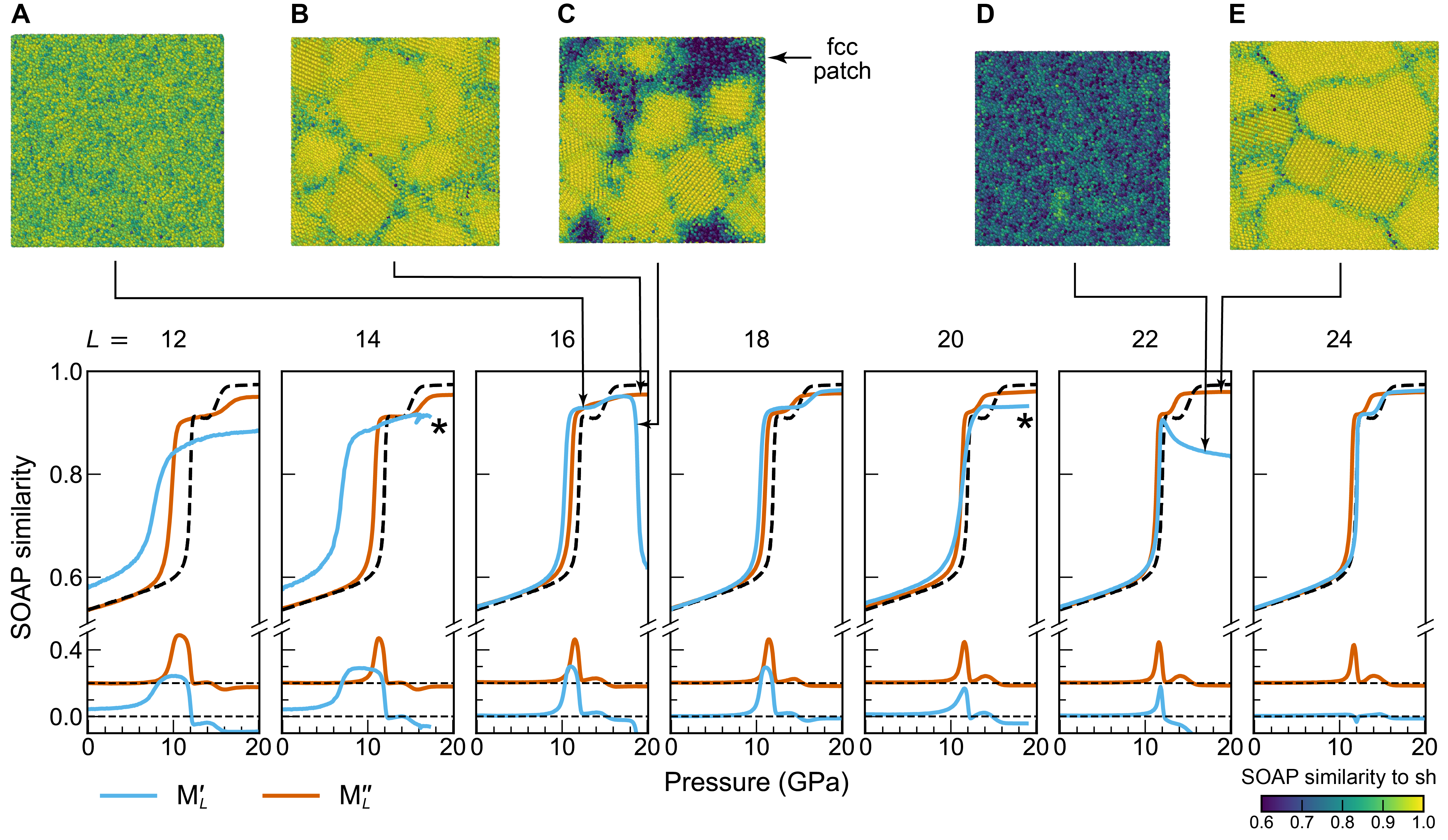}
    \caption{
    Performance of indirectly-learned potential models of increasing flexibility in 100,000-atom compression simulations of amorphous Si.
    We report the average SOAP similarity to pressure-adjusted sh Si during compression MD (cf.\ Fig.\ S1), benchmarking MTPs (solid) versus GAP-18 (dashed). The difference for each MTP to GAP-18 is plotted underneath the absolute values, with $\ind{M}_{L}$ offset for clarity. Representative snapshots during the run are displayed, colored according to atomistic SOAP similarity to sh (\textbf{A}--\textbf{E}). Points at which MD simulations fail (for $\direct{M}_{14}$ and $\direct{M}_{20}$) are marked with asterisks.
    }
    \label{fig:SOAP_difference}
\end{figure*}

\textit{Numerical errors.}---The accuracy of an ML potential is often characterized by an error versus its reference method \cite{Zuo2020}, typically evaluated on a test set not included during training. In this vein, Fig.\ \ref{fig:rmse} shows the mean absolute error on different test sets for a series of $\direct{M}$ and $\ind{M}\,$ models of increasing complexity and cost.
On the one hand, we test on random-structure-searching (RSS) configurations which are different from the fitting database; on the other hand, we test on snapshots from small-scale compression MD simulations, directly mirroring the physical situation that is to be ``indirectly learned''. 
 
For both test sets, Fig.\ \ref{fig:rmse} provides encouraging results: comparing their predictions to \textbf{M0} (DFT), indirectly-learned MTP models tend to at least match, if not improve upon, the corresponding directly-learned (\textbf{D0}-trained) MTPs. 
Inspecting the force-component errors (lower panels in Fig.\ \ref{fig:rmse}) indicates that our indirectly-learned MTP models faithfully recover the predictions of \textbf{M1}---suggesting that the existing \textbf{M1} PES is smooth enough to be learned by a different fitting method. More generally, the error analysis suggests that improvements to \textbf{M1} can be expected to transfer to subsequent indirectly-learned models.

We emphasize that the tests in Fig.\ \ref{fig:rmse} are not designed, nor able, to assess the quality of MTP models in absolute terms. In previous studies, reference databases for MTP fitting have been built using active learning \cite{Podryabinkin2017, Podryabinkin2019}, and it would now be interesting to include such a gradually improved database in the construction of MTP models for high-pressure disordered Si. 

\textit{Physically-guided validation.}---While numerical errors provide some indication of performance, the correlation between both is unreliable enough to warrant the extensive validation of ML potentials based on physical behavior---for example, assessing whether structural features of a reference simulation are reproduced. Here we use the Smooth Overlap of Atomic Positions (SOAP) kernel \cite{Bartok2013} which has proven useful for analyzing local structure in amorphous Si \cite{Bernstein2019a, aSi_100k}. The computational speed of ML potentials, and of MTPs in particular, allows us to compare several different indirectly-learned potentials in large-scale (100,000-atom) simulations. For Si, this provides a more stringent test than small-scale structures, allowing us to assess the description of the nucleation of sh crystallites and their intervening grain boundaries, and to quantify this description using SOAP. 

\begin{table}[t]
\caption{Quality metrics for 100,000-atom LDA Si models produced by \SI{e11}{K s^{-1}} quench simulations: the proportion of 3- and 5-fold connected atoms ($N_{3}$ and $N_{5}$), the inverse height of the first sharp diffraction peak ($H^{-1}$) \cite{Xie2013}, and the mean ($\bar{\theta}$) and width ($\Delta \theta$) of the bond-angle distribution.
}\label{tab:LDA}
\begin{tabular}{l
S[table-column-width=1.0cm, round-mode=places, round-precision=2]
S[table-column-width=1.0cm, round-mode=places, round-precision=2]
S[table-column-width=1.5cm, round-mode=places, round-precision=3]
S[table-column-width=1.2cm, round-mode=places, round-precision=2]
S[table-column-width=0.8cm, round-mode=places, round-precision=2]
S[table-column-width=1.0cm, round-mode=places, round-precision=3]} \toprule
    \centering
     &   {$N_{3}$} & {$N_{5}$} & {$H^{-1}$} & {$\bar{\theta}$} & {$\Delta \theta$} & {$\Delta E$}  \\ 
    {Model} & {(\%)} & {(\%)} &  & {(deg)} & {(deg)} & {(\si{eV})}  \\ 
    \colrule \\[-1em]
    $\direct{M}_{16}$   & 0.565     &   2.366       & 0.602664  & 108.99287      & 10.8465  & 0.156473  \\
    $\ind{M}_{16}$      & 0.677     &   1.637       & 0.560067  & 109.14605      & 10.2342  & 0.151846  \\
    \colrule \\[-1em]
    $\direct{M}_{20}$   & 0.471     &   3.645       & 0.498157  & 108.92970      & 10.9086  & 0.146161  \\ 
    $\ind{M}_{20}$      & 0.691     &   1.329       & 0.503525  & 109.16124      &  9.9319  & 0.143802  \\ 
    \colrule \\[-1em]
    GAP-18              & 0.696     &   0.952       & 0.563539  & 109.17405      &  9.9876  & 0.145155  \\
    \botrule
\end{tabular}
\end{table}

Figure \ref{fig:SOAP_difference} benchmarks a series of MTPs, directly-learned (blue; $\direct{M}_{L}$) or indirectly-learned (red; $\ind{M}_{L}$), with increasing quality and cost. All compression simulations start from the same low-density amorphous (LDA) structure from Ref.\ \citenum{aSi_100k} and are run using LAMMPS \cite{Plimpton1995}. In all cases, we use the similarity to sh Si as quality measure, aiming to quantify how well a given model reproduces the behavior of GAP-18 (\textbf{M1}). Overall, indirectly-learned MTPs perform much more reliably at the compression test than those directly fitted (to \textbf{D0}), with every $\ind{M}_{L}$ model producing the expected sequence of phases \cite{aSi_100k}, viz.\ LDA $\longrightarrow$ very-high-density amorphous (VHDA) $\longrightarrow$ polycrystalline (pc) sh. In contrast, only 2 out of 5 the directly-learned models perform acceptably. 
Such comparisons are only feasible because \textbf{M1} is efficient enough to produce a reference simulation (albeit with considerable effort), and additionally because \textbf{M2} is efficient enough to readily produce many of such simulations.

\begin{figure*}[ht]
    \centering
    \includegraphics[width=15cm]{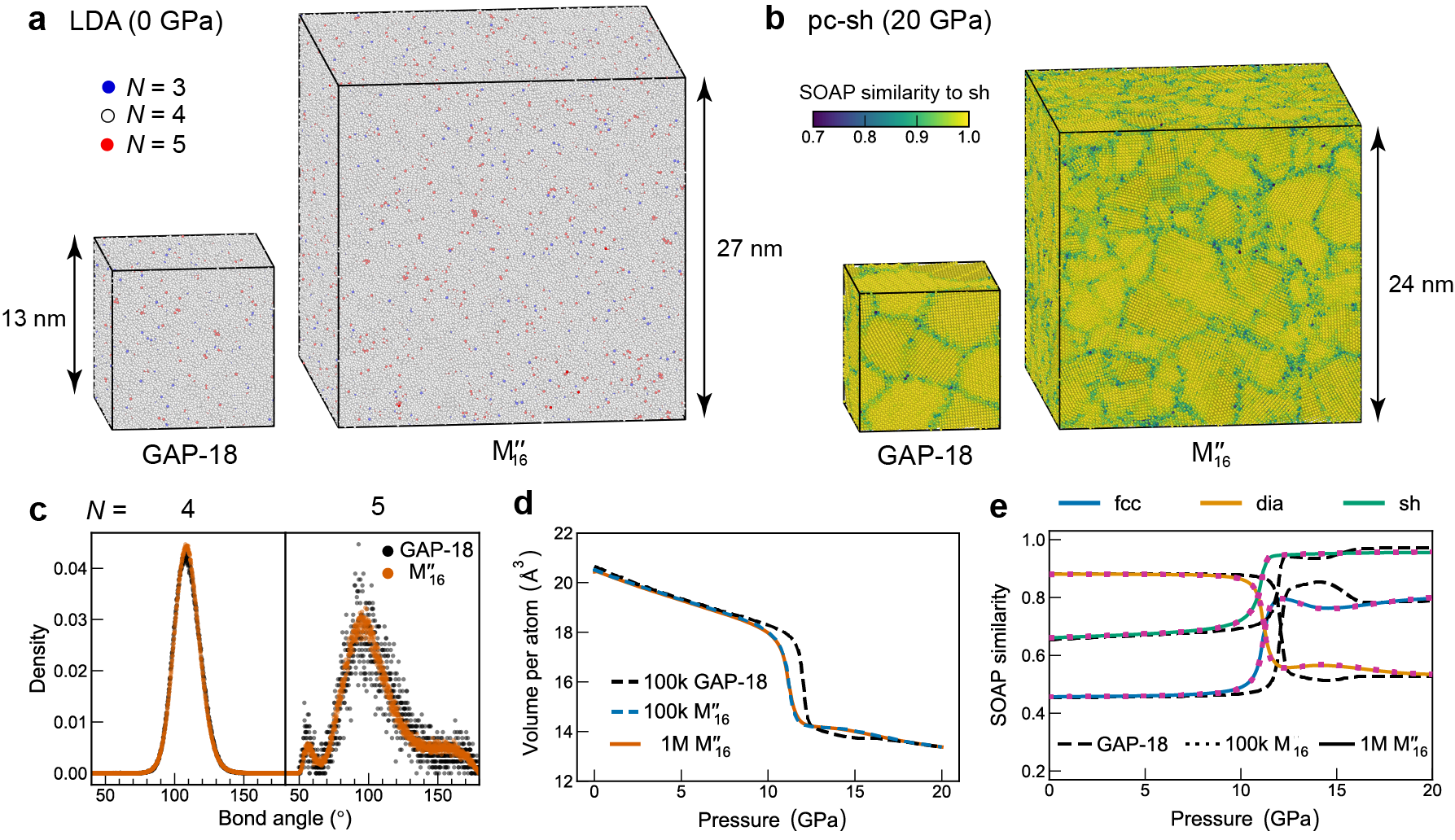}
    \caption{Million-atom MD simulations with an indirectly-learned potential. (a) Structural model of low-density amorphous (LDA) Si, obtained by quenching from the melt at \SI{e11}{Ks^{-1}} to \SI{500}{K} before relaxation. Color-coding indicates coordination numbers, $N$. 
    (b) Structural model of a polycrystalline (pc) simple-hexagonal (sh) phase produced by compression to \SI{20}{GPa}. Close inspection reveals a mixture of thick, amorphous interfaces and coherent symmetric-tilt grain boundaries. (c) Bond-angle distribution function of the LDA structure (cf.\ panel a) compared to GAP-18. The larger system-size permits a detailed analysis of $N = 5$ coordination defects. (d--e) Pressure--volume and pressure--similarity curves during compression of LDA $\longrightarrow$ VHDA $\longrightarrow$ pc-sh.
    Data for the previous GAP-18 simulation refer to a 100,000-atom system studied in Ref.\ \citenum{aSi_100k}.
    }
    \label{fig:1M}
\end{figure*}

We finally test our indirectly-learned potentials for a task for which they were not specifically trained (in \textbf{D1}): the production of LDA structural models via simulated quenching from the melt. Well-established quantitative measures are available \cite{AttaFynn2018,aSi_structures_JPCL,Limbu2020}
to measure how well our indirectly-learned models reproduce the behavior of GAP-18 (Table \ref{tab:LDA}).
We run quenching simulations with 100,000 atoms, as in Ref.\ \citenum{aSi_100k}, now comparing two representative levels of MTP fitting.
In both cases, the indirectly-learned model outperforms a directly-learned one at the same level. Both $\ind{M}$ models come close to reproducing the defect count, with that of 3-fold connected atoms being almost identical (0.68\%, 0.69\%, and 0.70\% with $\ind{M}_{16}$, $\ind{M}_{20}$, and GAP-18, respectively), and that of 5-fold connected ones being somewhat more varied. There is not, however, a unambiguous advantage of $\ind{M}_{20}$ over $\ind{M}_{16}$: specifically, the inverse height of the first sharp diffraction peak, $H^{-1}$, is much better described by the latter. This underlines the subtleties in assessing the behavior of ML potentials.

\textit{Application.}---To demonstrate its usefulness in practice, we employ our approach to generate million-atom structural models of LDA and pc-sh Si. 
The nanoscale grain structure of pc-sh invites the use of larger-scale models for more extended studies of the grain-boundary region and the size and orientation of crystallites. 
The efficiency of MTPs permits fast simulations on long timescales without having to compromise on system size. 
Figure \ref{fig:1M} shows results of a melt-quench (Fig.\ \ref{fig:1M}a) and compression MD (Fig.\ \ref{fig:1M}b) simulation of a system of 1M Si atoms, using the $\ind{M}_{16}$ model \footnote{For recent, large-scale atomistic simulations with ML potentials, see, e.g., Ref.\ \citenum{Smith2021}, or: W. Jia, H. Wang, M. Chen, D. Lu, L. Lin, R. Car, W. E, and L. Zhang, in \textit{SC '20: Proceedings of the International Conference for High Performance Computing, Networking, Storage and Analysis}, 5 (2020).
}. 

The LDA structure, visualized in Fig.\ \ref{fig:1M}a, has a bond-angle distribution closely similar to that for the 100k-atom GAP-18-generated structure (Fig.\ \ref{fig:1M}c). However, a more detailed analysis of the environments of the $N = 5$ atoms (``coordination defects'') is possible with larger system size, as evident from the much lower scatter in the results shown in Fig.\ \ref{fig:1M}c. The histogram bin widths are  \SI{0.25}{\degree} for both system sizes for fair comparison, sufficiently narrow to permit analysis of small deviations away from the ideal tetrahedral angle. The wide distribution of bond angles in $N = 5$ atoms is consistent with the relatively wide spread in local structure, as characterized in Ref.\ \citenum{Bernstein2019a} for much smaller simulation cells.

Upon compression, the expected pressure-induced collapse of the largely fourfold-connected structure into a VHDA phase \cite{Durandurdu2001, aSi_100k} is reproduced by the indirectly-learned potential, the onset occurring at only slightly lower pressure compared to the reference GAP-18 simulation. There are no discernible differences between the volume--pressure curves of 100k-atom and 1M-atom simulations carried out using the same indirectly-learned model (Fig.\ \ref{fig:1M}d), making system-size effects for VHDA {\em formation} unlikely. However, system-size effects in the nucleation of sh crystallites are clearly still important even in 100k-atom systems. On the top face of the GAP-18 structure in Fig.\ \ref{fig:1M}b, a crystallite extends all across the cell and hence is infinite in that direction under periodic boundary conditions. The 1M-atom structure generated by $\ind{M}_{16}$ has similar grain sizes, but the larger cell ensures that no grain approaches the length of the cell itself.

The differences in the volume--pressure curves between GAP-18 and $\ind{M}_{16}$ (Fig.\ \ref{fig:1M}d) are of a  similar magnitude to those between GAPs trained using different reference methods in Ref.\ \citenum{aSi_100k}. The crystallization and volume of the pc-sh phase are well reproduced, as is the behavior as characterized using SOAP (Fig.\ \ref{fig:1M}e). 
Figure \ref{fig:1M} therefore suggests that our indirectly-learned $\ind{M}_{16}$ model {\em can}, to a large extent, match predictions of GAP-18.
This way, more detailed analyses become possible---for example, of the grain structure in pc-sh Si, as well as a closer study of the LDA phase which is the subject of ongoing work.

In future studies, we envisage using more diverse combinations of fitting frameworks in a similar vein, e.g., neural networks \cite{Behler2007} and SNAP \cite{Thompson2015}, and automating this model selection. In addition, we propose to incorporate ensembles of base models, including using different levels of theory, to further augment the capabilities of an indirectly-learned model. We hope that such developments will help to make quantum-mechanically accurate materials simulations with millions of atoms more commonplace in the years ahead.

\normalsize

\begin{acknowledgements}
J.D.M. acknowledges funding from the EPSRC Centre for Doctoral Training in Inorganic Chemistry for Future Manufacturing (OxICFM), EP/S023828/1.
The authors acknowledge the use of the University of Oxford Advanced Research Computing (ARC) facility in carrying out this work (http://dx.doi.org/10.5281/zenodo.22558), as well as support from the John Fell Oxford University Press (OUP) Research Fund.
We thank S. Batzner, D. Holzm\"u{}ller, S. Hoyer, and A. Shapeev for helpful comments on the manuscript.
\end{acknowledgements}

\cleardoublepage 
\newpage
\onecolumngrid

\begin{center}

{\fontsize{11.5}{14}\selectfont \textbf{Supplemental Material for}\\
``Indirect Learning of Interatomic Potentials for Accelerated Materials Simulations''\\[4mm]}

{\fontsize{10}{14}\selectfont Joe D. Morrow and Volker L. Deringer*\\[1mm]}
{\fontsize{9}{11}\selectfont \textit{Department of Chemistry, Inorganic Chemistry Laboratory,\\ University of Oxford, Oxford OX1 3QR, United Kingdom}\\[6mm]}

\end{center}

\setcounter{table}{0}
\setcounter{figure}{0}
\renewcommand{\thefigure}{S\arabic{figure}}
\renewcommand{\thetable}{S\arabic{table}}

\setcounter{equation}{0}
\renewcommand{\theequation}{S\arabic{equation}}

\setcounter{page}{1}
\renewcommand{\thepage}{S\arabic{page}}
\setcounter{section}{0}
\renewcommand{\thesection}{S\arabic{section}}
\setcounter{subsection}{0}
\renewcommand{\thesubsection}{S\arabic{section}.\arabic{subsection}}

\twocolumngrid

\section*{SOAP similarity analysis for physically-guided validation} 
The Smooth Overlap of Atomic Positions (SOAP) descriptor is a fingerprint originally developed in the context of ML potentials \cite{Bartok2013} and widely employed in GAP fitting \cite{chemrev}; however, SOAP has has also been used for classifying and understanding complex structures across a wide range of chemical systems [\citenum{Musil2021}, S1--S2]. 
Briefly, the SOAP formalism encodes the local environment of atom $i$ as a neighbor density $\rho^{(i)}(\mathbf{r})$ by placing a 3D Gaussian function of width $\sigma_{\mathrm{at}}$ on $i$ and each of its neighbors up to a cutoff. A similarity measure $k^{\mathrm{SOAP}}(i, j)$ between two such environments is given by evaluating the overlap integral of their respective $\rho(\mathbf{r})$ and averaging over all possible rotations $\hat{R}$ of one of the local environments. 

\begin{figure}[t]
\includegraphics[width=7.5cm]{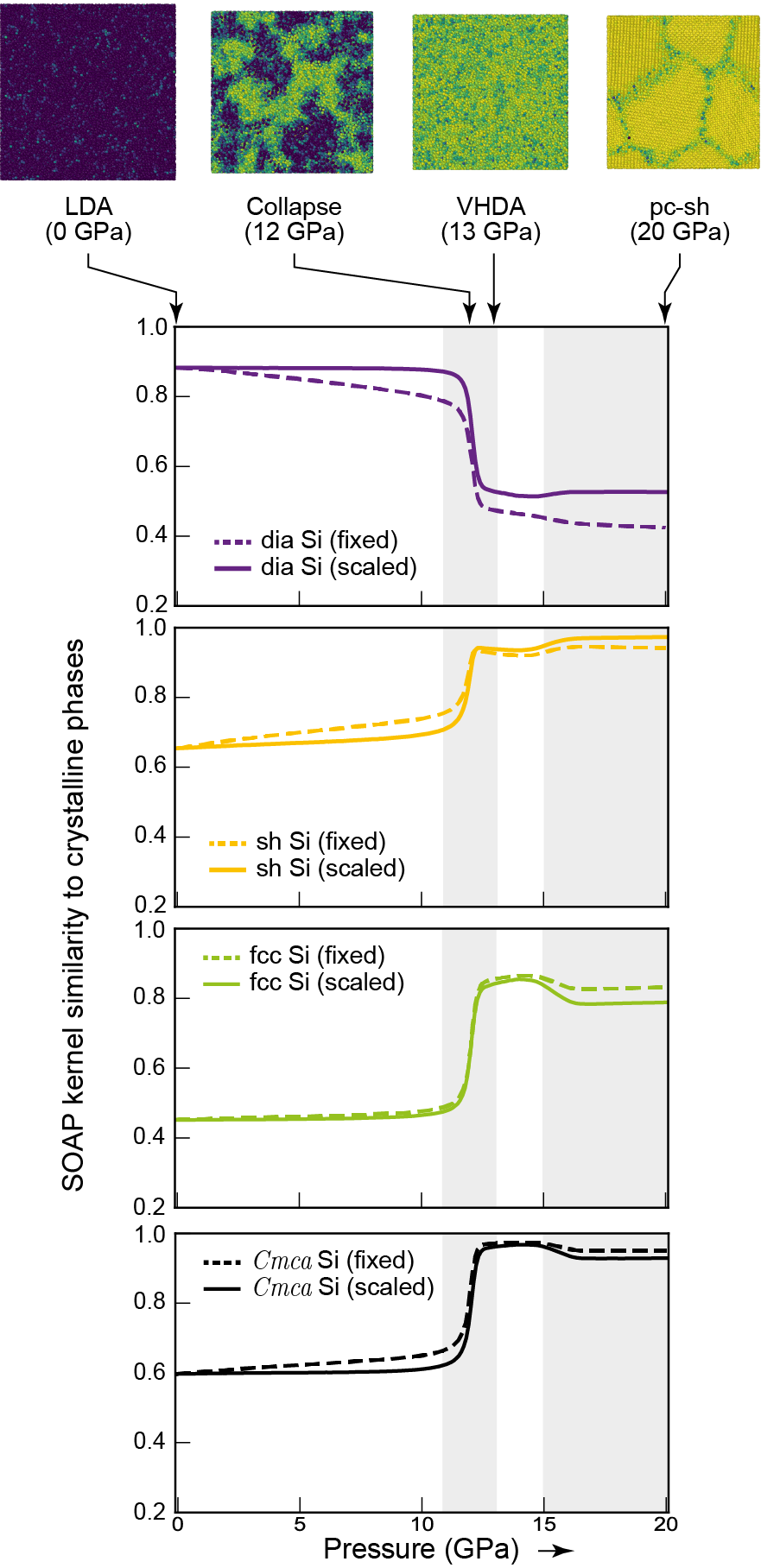}
\caption{A structural benchmark for high-pressure disordered Si. We evaluate the average SOAP similarity of the compression trajectory of Ref.\ \citenum{aSi_100k} to selected crystalline structures. Dashed lines are for reference crystals with a fixed lattice optimised at \SI{0}{GPa} with DFT. Solid lines are similarities to crystals with their lattice parameters optimised as a function of pressure. The structure images show representative snapshots, color-coded according to SOAP similarity to the sh crystalline phase, with lighter colors indicating higher similarity (drawn with data from Ref.\ \citenum{aSi_100k}).
}
\label{fig:SOAP_fix_v_adjust}
\end{figure}

Here we use the configuration-averaged SOAP kernel (cf.\ Ref.\ S2) to measure the similarity of MD frames to crystal structures adopted by Si, which provides a physical interpretation of the average structure in a large simulation system. We are interested in the structural changes that occur with increasing pressure beyond the trivial isotropic contraction of bond lengths. To subtract this effect from that of the SOAP similarity, we compare our simulation trajectory not only to ambient-pressure crystals (``fixed'' in Fig.\ \ref{fig:SOAP_fix_v_adjust}), but also to crystals with lattice parameters optimized as a function of pressure. More precisely, we optimize the lattice parameters at \SI{1}{GPa} intervals with DFT and interpolate between them when calculating the similarity so that the reference crystal is under the same pressure as is measured over each \si{ps} of MD simulation time (``scaled'' in Fig.\ \ref{fig:SOAP_fix_v_adjust}). 

Figure \ref{fig:SOAP_fix_v_adjust} shows the evolution of the SOAP similarity with pressure during the compression simulation reported in Ref.\ \citenum{aSi_100k} (carried out for a 100,000-atom system using GAP-18) which we here use as reference, assessing the per-atom similarity to crystalline phases in two different ways: either fixing them to their ambient-pressure structures (dashed lines), or allowing for freely varying lattice parameters with pressure (solid lines). The atomic environments in the quenched LDA phase at \SI{0}{GPa} have a rather close resemblance to those in the cubic diamond (\textbf{dia}) structure, with both featuring tetrahedral Si environments. Up to about \SI{12}{GPa}, the similarity to pressure-adjusted \textbf{dia} remains practically constant. 

\FloatBarrier 

\begin{figure}[t]
    \centering
    \includegraphics[width=\linewidth]{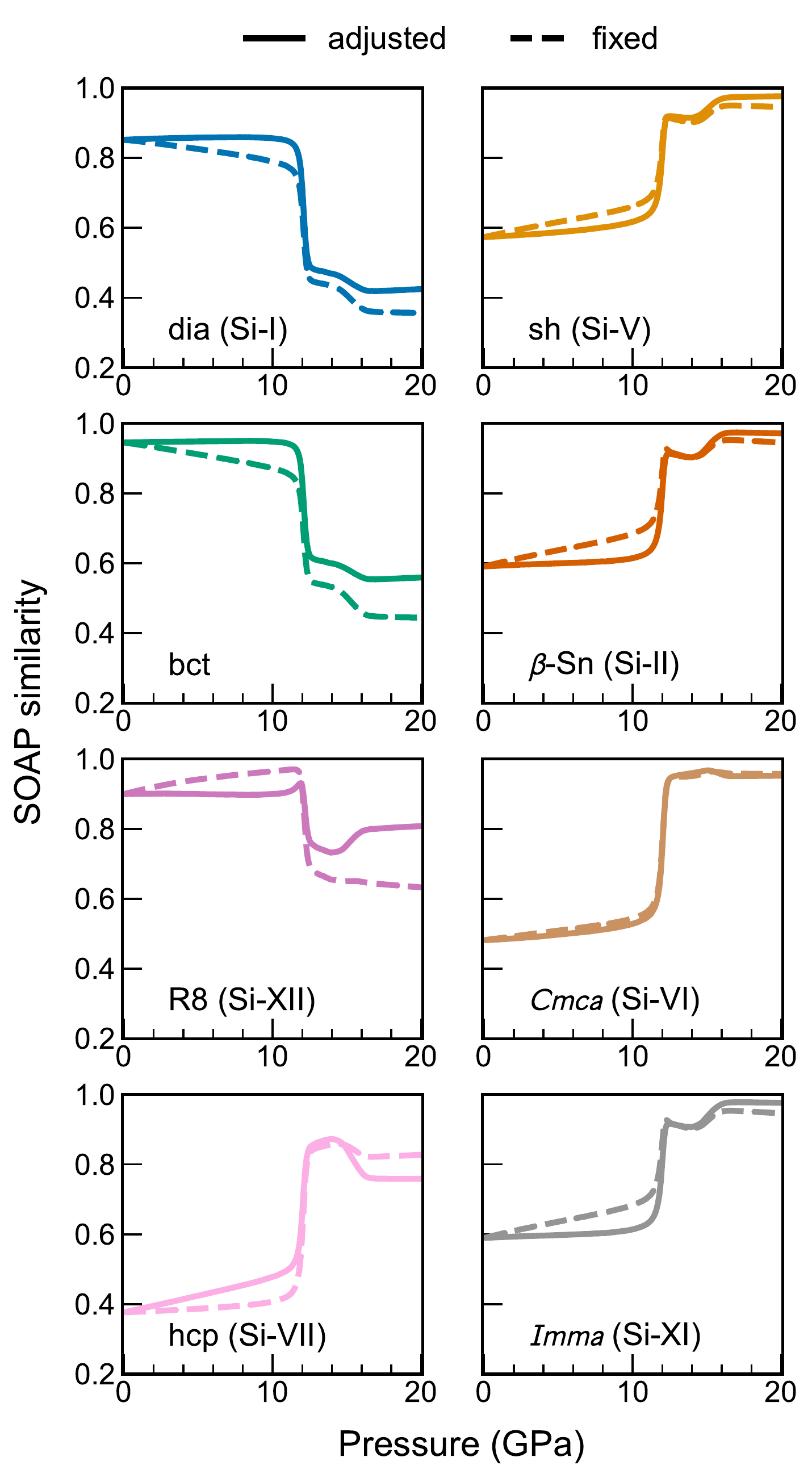}
    \caption{Extension of Fig.\ \ref{fig:SOAP_fix_v_adjust} to now include further crystal structures: comparing the average atomic environment during compression MD to reference structures. Dashed lines are for reference crystals with fixed structural parameters as optimized using DFT at \SI{0}{GPa}. Solid lines indicate SOAP similarities to crystals with their lattice parameters optimized using DFT as a function of pressure. Rhombohedral phase R8 results from decompressing $\beta$-Sn, containing 5 and 6-membered rings of tetrahedrally-coordinated Si (see Ref.\ S3).
    bct is a hypothetical four-fold coordinated structure predicted to be stable under tensile stress [S4]. The metallic  phases commonly labeled with their orthorhombic spacegroups, $Imma$ and $Cmca$, appear with increasing pressure respectively between $\beta$-Sn and sh and between sh and hexagonally close-packed (hcp) phases. 
    Anzellini {\em et al.} [S5] provide an introduction to the numerous Si phases and their nomenclature.
    }
    \label{fig:SI_SOAP_extended}
\end{figure}

More detailed inspection of the MD trajectory confirms the interpretation that the LDA phase does not change substantially beyond an isotropic contraction until it collapses to a very-high-density amorphous phase (VHDA) around \SI{12}{GPa} \cite{aSi_100k}, seen as a sudden drop in the \textbf{dia} similarity. The VHDA phase has the closest similarity among tested crystals to the high-pressure Si-VI phase with space group \textit{Cmca}, which appears experimentally upon compression to about \SI{40}{GPa}, after the formation of sh and before hexagonal close-packed (hcp) Si. At about 14 to \SI{16}{GPa}, the nucleation of sh crystallites begins \cite{aSi_100k}, so the similarity to \textit{Cmca} decays slightly and similarity to sh increases. At this point, characterizing inhomogeneous polycrystalline sh (pc-sh) with any single variable becomes fraught, which also explains why the \textit{Cmca} similarity remains so high (along with $\beta$-Sn; see Fig.\ \ref{fig:SI_SOAP_extended}). The interfaces between sh grains, which are more similar to \textit{Cmca} and $\beta$-Sn than they are to sh, occupy a reasonable fraction of the simulation-cell volume, hence contribute appreciably to the average similarity. Nevertheless, the analysis in Fig.\ \ref{fig:SOAP_fix_v_adjust} summarizes the structural changes during compression in a comprehensive way.

\begin{table}[t]
\caption{General hyperparameters used for SOAP-based structural analysis, carried out using the QUIP/GAP code as interfaced to the \texttt{quippy} Python package (see https://github.com/libAtoms/QUIP). SOAP vectors were configuration-averaged ($\texttt{average}=\mathrm{T}$ in \texttt{quippy}) for each MD frame before their transformation to the power spectrum and then kernel via a dot product with the reference crystal. The dot product was raised to a power of $\zeta=4$.}\label{tab:soap_hypers}
\begin{tabular}{l >{\centering\arraybackslash}m{10em}}
\toprule
    \centering
   $\sigma_{\mathrm{at}}$ & \SI{0.5}{\angstrom} \\
   $R_{\mathrm{cut}}$ & \SI{5.0}{\angstrom} \\
    $l_\text{max}$ & 6 \\
    $n_\text{max}$ & 10 \\
    \texttt{cutoff\_transition\_width} & \SI{2.0}{\angstrom} \\
    $\zeta $ & 4 \\
\botrule
\end{tabular}
\end{table}

\FloatBarrier

\section*{Melt--quench simulations}

\begin{figure}[b]
    \centering
    \includegraphics[width=8.25cm]{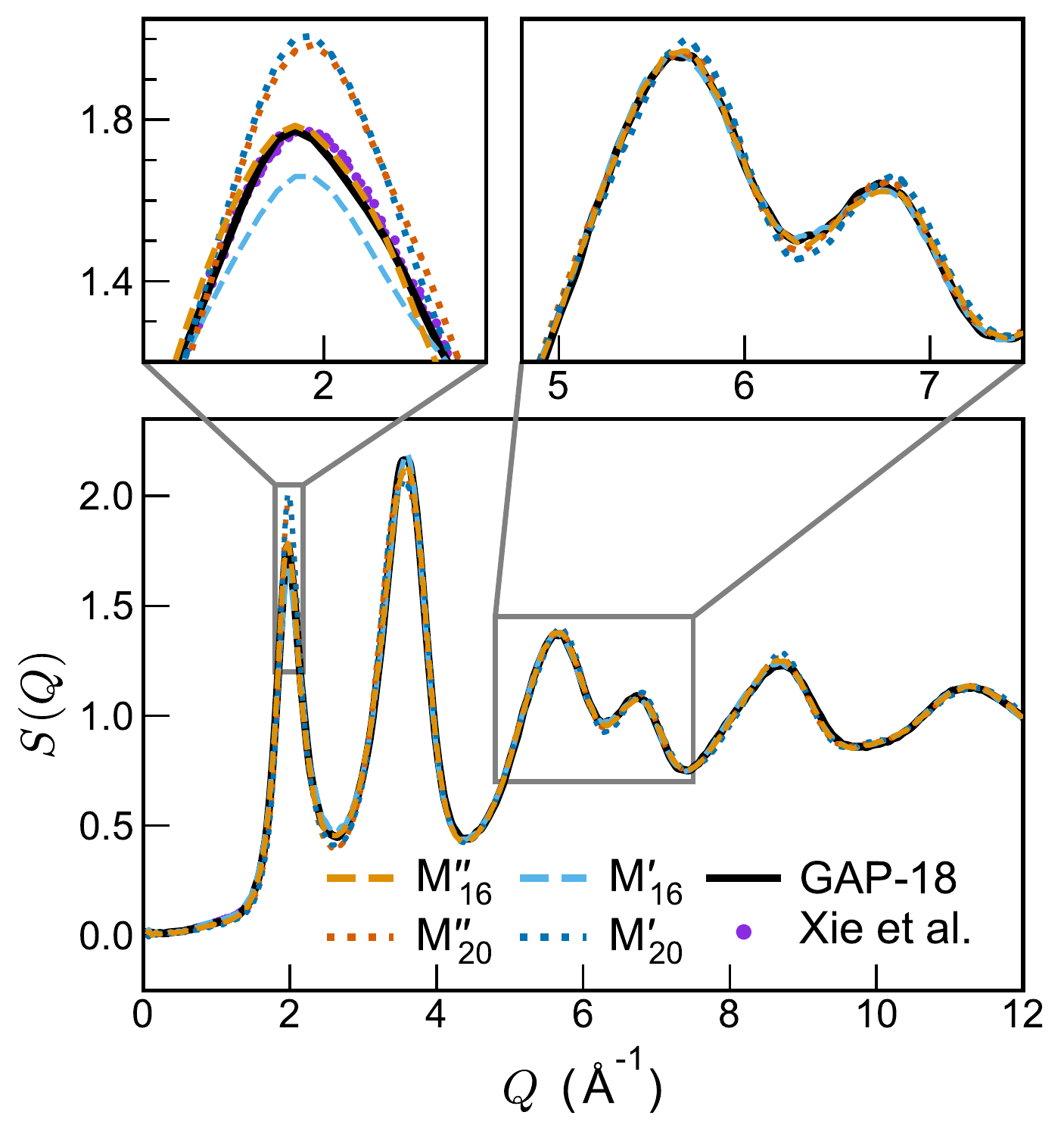}
    \caption{Variable-rate melt quenches as a benchmark for the behavior of ML potentials for Si. The figure compares \textbf{D0}-trained and indirectly-learned MTPs to \textbf{M1} (upper panels) and experiment (lower panel); the simulations follow the protocol of Ref.\ \citenum{aSi_structures_JPCL}.
    Although the structure factor, $S(Q)$, as predicted by $\ind{M}_{16}$, agrees almost perfectly with the GAP-18 result, the much larger FSDP predicted by $\ind{M}_{20}$ suggests this may in fact be serendipitous. It is nevertheless encouraging that the other peaks are also well-matched by the $\ind{M}_{16}$ curve. The increased ordering implied by the higher FSDP of the $\ind{M}_{20}$ curve does not obviously manifest as crystalline grains, nor is any peak splitting characteristic of paracrystalline ordering clear in the other diffraction peaks.}
    \label{fig:Sq_M16_M20_100k_v_GAP100k}
\end{figure}

\begin{figure}[t]
    \centering
    \includegraphics[width=\linewidth]{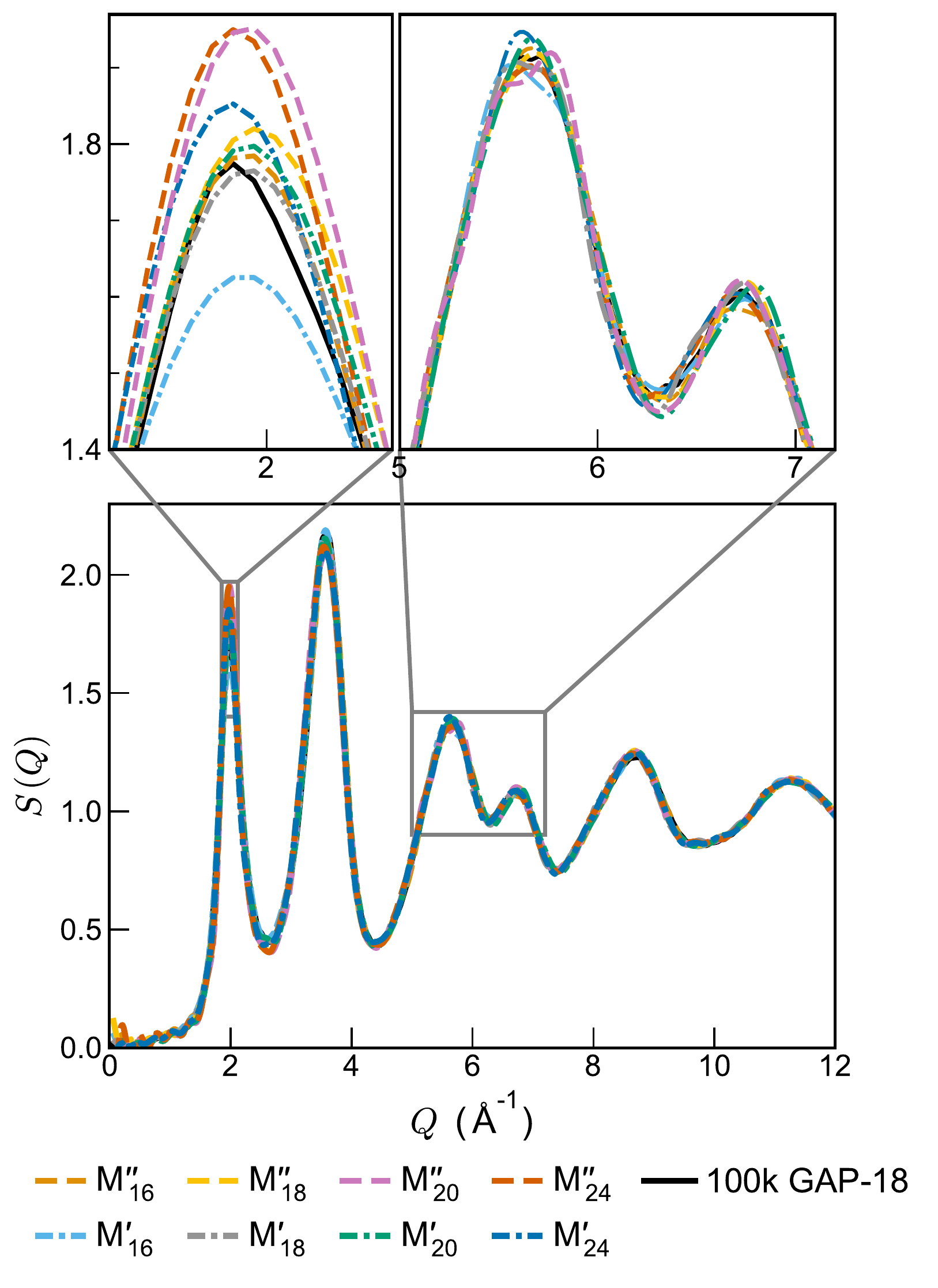} 
    \caption{Static structure factors of 4096-atom low-density amorphous Si structures, produced via the same variable-temperature melt-quench protocol as for Figure \ref{fig:Sq_M16_M20_100k_v_GAP100k}, and using directly and indirectly fitted MTPs covering a broader range of $\text{lev}_{\text{max}}$ values. Some of the variability in the height of the FSDP can be explained by the stochastic nature of grain formation during cooling in combination with the relatively small system size (the height varies from simulation to simulation for the same MTP and same protocol: e.g., compare $\direct{M_{20}}$ in this figure and Fig.\ \ref{fig:Sq_M16_M20_100k_v_GAP100k}). However, the more flexible indirectly-learned MTPs do seem to induce some over-ordering. More detailed studies using indirectly-learned MTPs specifically trained for melt-quench simulations will be the subject of future work.
    }
    \label{fig:Sq_SI}
\end{figure}

The static structure factor, $S(Q)$, is a particularly useful quantity because it can be experimentally observed, and high-quality reference data for amorphous Si are available (e.g., Refs.\ \citenum{Xie2013} and S6).
Figure \ref{fig:Sq_M16_M20_100k_v_GAP100k} shows the structure factor of an LDA model produced using the variable quench-rate protocol of Ref.\ \citenum{aSi_structures_JPCL}. There is almost perfect agreement between $\ind{M}_{16}$ and GAP-18, which also matches the experimental data from Ref.\ \citenum{Xie2013} closely. Smaller-scale simulations suggest that levels 16 and 20 span most of the variation in $H^{-1}$ among the MTPs studied here (Fig.\ \ref{fig:Sq_SI}), but system sizes of the order of 100k atoms are required for repeatable results due to the stochastic nature of the simulations.  

We limit ourselves to two levels because of the expense associated with the long timescales of \SI{e11}{K.s^{-1}} melt-quenching, but note that other indirectly-learned MTPs tested tend to overestimate the FSDP height, with $\ind{M}_{12}$ and $\ind{M}_{14}$ producing regions of diamond-like crystallinity (as do $\direct{M}_{12}$ and $\direct{M}_{14}$); this requires further work in a future study, and we focus on the level 16 and 20 models herein. The strong performance of $\ind{M}_{16}$ supports our choice to use it for a large-scale proof-of-concept (Fig.\ \ref{fig:1M} in the main text).
\FloatBarrier 
\cleardoublepage

\section*{Computational details}

\subsection*{Moment tensor potential fitting}
MTP models have been applied to a range of research questions, including phase diagrams \cite{Rosenbrock2021}, reaction dynamics of small molecules [S7], and lithium ion conduction [S8]. Full details of their construction are given in the original work by Shapeev and co-workers \cite{Shapeev2016,Novikov2021}, which we summarize briefly here. The descriptor of a local environment is defined as the moment tensor,
\begin{equation}
    M_{\mu,\nu}(\mathbf{n}_i) = \sum\limits_j f_{\mu}(|\mathbf{r}_{ij}|) \; \underbrace{\mathbf{r}_{ij} \otimes ... \otimes \mathbf{r}_{ij}}_{\nu \; \mathrm{times}},
\end{equation}
where $\mathbf{n}_i$ denotes the positions of the $i^{\mathrm{th}}$ atom and all neighbors $j$ up to a cutoff, $R_\mathrm{cut}$, which we set to \SI{5}{\angstrom} consistent with that of the descriptor used for GAP-18 \cite{Bartok2018}. The radial part $f_{\mu}$ depends only on the position vectors $\mathbf{r}_{ij}$ of each $j$-th atom from the $i$-th one, and the angular information is encoded by tensors of rank $\nu$ produced from the repeated outer product of $\mathbf{r}_{ij}$ vectors. These $M_{\mu,\nu}$ are contracted to give the basis functions $B_{\alpha}$. Training involves finding the regression coefficients $\xi_{\alpha}$ that minimize a loss function based on the errors in predicted energies, 
\begin{equation}
    E_{\rm MTP} = \sum_{i} \varepsilon_{i} \equiv \sum_i \sum_{\alpha} \xi_{\alpha}B_{\alpha}( \mathbf{n}_i ),
\end{equation}
as well as forces and stresses, which can be expressed in terms of the first and second derivatives of $E_{\rm MTP}$ with respect to $\mathbf{r}_{ij}$. The number of basis functions defining the particular functional form of an MTP (and hence its flexibility and computational efficiency) grows exponentially with the important hyperparameter $\mathrm{lev_{max}}$, where the level (``lev'') is defined as
\begin{subequations}
    \begin{align}
    \mathrm{lev}(M_{\mu,\nu}) &= 2 + 4\mu + \nu\\
    \mathrm{lev}(M_{\mu,\nu}\!:\!M_{\eta,\lambda}) &= \mathrm{lev}(M_{\mu,\nu}) + \mathrm{lev}(M_{\eta,\lambda})
    \end{align}
\end{subequations}
and ``$:$'' represents an appropriate tensor contraction operation.
All possible $M_{\mu,\nu}$, and contractions of one or more $M_{\mu,\nu}$, are included as degrees of freedom in the functional form of an MTP, provided the resulting basis function satisfies $\mathrm{lev}(B_{\alpha})\leq \mathrm{lev_{max}}$. 
In the present work, we investigated the range $\mathrm{lev_{max}}\!=\!$ 12--24. For brevity, we denote $\mathrm{lev_{max}}$ as ``$L$'' in the main text.

We found that the performance of MTPs was quite sensitive to the relative weights of energies, forces, and stresses in the fit. After some testing, we settled on $10:1:0.01$ respectively. Our chosen force weight is a factor of 10 larger than the commonly-used default value, as we found both direct and indirectly-learned MTPs suffered from unstable MD and poor predictions of the physical behavior under pressure using the default force weight. 
Full details of all remaining MTP hyperparameter choices are given in Table \ref{tab:mtp_hypers}.

\begin{table}[t]
\caption{General hyperparameters used for all MTPs trained in this work. The default \texttt{radial\_basis\_size} for each value of $\text{lev}_{\text{max}}$ was used throughout, as was the case for any parameters not listed below.}\label{tab:mtp_hypers}
\begin{tabular}{l >{\centering\arraybackslash}m{10em}}
\toprule
    \centering
    \texttt{min\_dist} &  \SI{1.5}{\angstrom} \\
    \texttt{max\_dist} &  \SI{5.0}{\angstrom} \\
    \texttt{radial\_basis\_type} & \texttt{RBChebyshev} \\
    \texttt{energy-weight} & 1.0 \\
    \texttt{force-weight} & 0.1 \\
    \texttt{stress-weight} & 0.001 \\
    MTP version & 1.1.0 \\
\botrule \\[4mm]
\end{tabular}
\end{table}

\begin{figure}[t]
\includegraphics[width=\linewidth]{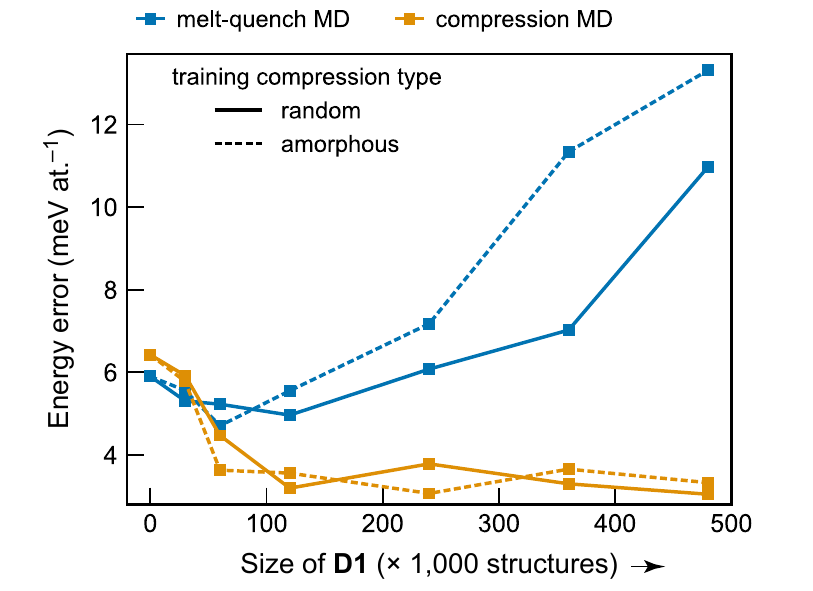}
\caption{
Mean absolute error of predicted energies versus quantity of compression MD training data of two types: starting separately from LDA structures (dashed lines) and randomly placed hard spheres (solid lines). 3 $\ind{M}_{16}$ MTPs with the same settings were trained for each database and we report their average errors in energy prediction vs. GAP-18 on two independent validation sets: structures derived from melt-quench MD (blue) and compression MD (yellow). The median standard deviation in error across identically-trained MTPs, $\sigma_{\Delta E}/|\Delta E|$, is in the region of \SIrange{10}{45}{\%}. MAEs show improvement with the addition of similar data to the training set, which is converged with respect to database size by \SI{240e3} atomic environments.  
The errors for melt-quench MD begin to degrade with the increased proportion of compression MD included during training between \SIrange{120e3}{240e3} environments. Including even more environments leads to an unstable potential for MD simulations.
}
\label{fig:rmse_data}
\end{figure}
\FloatBarrier
\clearpage 

\subsection*{\textbf{D1} database}
The reference configurations for the indirect learning step were obtained with an MD simulation protocol of \SI{50}{ps} thermalization of a random initial structure at \SI{500}{K}, followed by a constant pressure ramp to \SI{20}{GPa} over \SI{200}{ps} (as used in Ref.\ \citenum{aSi_100k}).
In Fig.\ \ref{fig:rmse_data}, we study errors for compression MD with respect to quantity of compression training data, which we find to converge by around \SI{2e5} atomic environments. Meanwhile, the errors for melt-quench MD start to {\em diverge} at this point (becoming worse with larger \textbf{D1}), suggesting that over-weighting environments encountered during compression MD comes at the expense of performance in other regions of configuration space. Based on these results, and our observations of unstable MD for compression data sets much greater than \SI{2e5} environments, we found potentials trained on \SI{2.4e5} environments from random starting points to be suitable and therefore use a database of this size for validation.

\subsection*{Test configurations}
The ``RSS'' (random structure searching) test set, for which results are shown in Fig.\ \ref{fig:rmse}a in the main text, consists of 100 structures produced using the {\tt buildcell} code of the Ab Initio Random Structure Searching (AIRSS) software \cite{Pickard2006, Pickard2011} and optimized with \textbf{M1} at pressures in the range \SIrange{0}{32}{GPa}. The initial geometries of the random structures were constrained to have a minimum interatomic separation of \SI{2}{\angstrom}, a volume of \SI{20(5)}{\angstrom^3} per atom, \SIrange{1}{4} random symmetry operations, and \SIrange{1}{64} atoms in the unit cell. The corresponding {\tt buildcell} settings were \texttt{NATOM=1-16, VARVOL=20, MINSEP=2.0,} \texttt{ SYMMOPS=1-4}. Each structure was optimized at a random pressure taken from the (positive) exponential distribution centered at \SI{0}{GPa} with width \SI{5}{GPa}. See Ref.\ \citenum{Pickard2011} for details of the AIRSS framework.

The ``compression MD'' set (Fig.\ \ref{fig:rmse}b) consists of structures from 64-atom compression runs of LDA Si (snapshots taken between 5 and 20 GPa). Both test sets were evaluated in single-point DFT computations with settings corresponding to Ref.\ \citenum{Bartok2018} (cf.\ https://doi.org/10.5281/zenodo.1250555), using the PW91 functional \cite{Perdew1992}, a 250 eV energy cut-off (with automatic finite basis set correction), 0.05 eV Fermi--Dirac smearing, and an on-the-fly pseudopotential, as implemented in CASTEP \cite{Clark2005}.

\subsection*{Computational cost}
Timing data (used to construct the horizontal axis in Fig.\ \ref{fig:rmse} in the main text) were acquired by averaging across 9 MD runs of 5,000 timesteps with 10,000 atoms, parallelized across 48 CPU cores using LAMMPS. The protocol, initial structure, and random seed used were the same for each potential.

The 1-million atom compression simulation was parallelized across 768 CPU cores, requiring a total wall-time of 15.5 hours for \SI{0.25}{ns} of simulation time: \num{11.9e3} CPU hours, \num{2.50e5} timesteps, \SI{5.85e3}{\text{timesteps}.\text{atom}^{-1}.\text{core-second}^{-1}}, where forces are evaluated at each \SI{1}{fs} timestep. 
The 1-million atom melt-quench simulation was parallelized across 384 CPU cores, requiring a total wall-time of 195.6 hours for \SI{2.22}{ns} of simulation time: \num{75.1e3} CPU hours, \num{2.22e6} timesteps, \SI{8.2e3}{\text{timesteps}.\text{atom}^{-1}.\text{core-second}^{-1}}.

\section*{Supplementary references}
\footnotesize 

\renewcommand{\theenumi}{S\arabic{enumi}}
\begin{enumerate}
    \item S. De,  A. P. Bart\'o{}k,  G. Cs\'a{}nyi,   and M. Ceriotti, Phys. Chem. Chem. Phys. {\bf 18}, 13754 (2016).
    \item J. Mavra\v{c}i\'c, F. C. Mocanu, V. L. Deringer, G. Cs\'a{}nyi, and S. R. Elliott, J. Phys. Chem. Lett. {\bf 9}, 2985 (2018).
    \item J. Crain, G. J. Ackland, J. R. Maclean, R. O. Piltz, P. D. Hatton, and G. S. Pawley, Phys. Rev. B {\bf 50}, 13043(R) (1994).
    \item Y. Fujimoto, T. Koretsune, S. Saito, T. Miyake, and A. Oshiyama, New J. Phys. {\bf 10}, 083001 (2008).
    \item S. Anzellini, M. T. Wharmby, F. Miozzi, A. Kleppe, D. Daisenberger, and H. Wilhelm, Sci. Rep. {\bf 9}, 15537 (2019).
    \item K. Laaziri, S. Kycia, S. Roorda, M. Chicoine, J. L. Robertson,  J. Wang,   and S. C. Moss, Phys. Rev. B {\bf 60}, 13520 (1999).
    \item I. S. Novikov, Y. V. Suleimanov,  and A. V. Shapeev, Phys. Chem. Chem. Phys. {\bf 20}, 29503 (2018).
    \item C. Wang, K. Aoyagi, P. Wisesa,   and T. Mueller, Chem. Mater. {\bf 32}, 3741 (2020).
\end{enumerate}


\begin{thebibliography}{54}%
\makeatletter
\providecommand \@ifxundefined [1]{%
 \@ifx{#1\undefined}
}%
\providecommand \@ifnum [1]{%
 \ifnum #1\expandafter \@firstoftwo
 \else \expandafter \@secondoftwo
 \fi
}%
\providecommand \@ifx [1]{%
 \ifx #1\expandafter \@firstoftwo
 \else \expandafter \@secondoftwo
 \fi
}%
\providecommand \natexlab [1]{#1}%
\providecommand \enquote  [1]{``#1''}%
\providecommand \bibnamefont  [1]{#1}%
\providecommand \bibfnamefont [1]{#1}%
\providecommand \citenamefont [1]{#1}%
\providecommand \href@noop [0]{\@secondoftwo}%
\providecommand \href [0]{\begingroup \@sanitize@url \@href}%
\providecommand \@href[1]{\@@startlink{#1}\@@href}%
\providecommand \@@href[1]{\endgroup#1\@@endlink}%
\providecommand \@sanitize@url [0]{\catcode `\\12\catcode `\$12\catcode
  `\&12\catcode `\#12\catcode `\^12\catcode `\_12\catcode `\%12\relax}%
\providecommand \@@startlink[1]{}%
\providecommand \@@endlink[0]{}%
\providecommand \url  [0]{\begingroup\@sanitize@url \@url }%
\providecommand \@url [1]{\endgroup\@href {#1}{\urlprefix }}%
\providecommand \urlprefix  [0]{URL }%
\providecommand \Eprint [0]{\href }%
\providecommand \doibase [0]{http://dx.doi.org/}%
\providecommand \selectlanguage [0]{\@gobble}%
\providecommand \bibinfo  [0]{\@secondoftwo}%
\providecommand \bibfield  [0]{\@secondoftwo}%
\providecommand \translation [1]{[#1]}%
\providecommand \BibitemOpen [0]{}%
\providecommand \bibitemStop [0]{}%
\providecommand \bibitemNoStop [0]{.\EOS\space}%
\providecommand \EOS [0]{\spacefactor3000\relax}%
\providecommand \BibitemShut  [1]{\csname bibitem#1\endcsname}%
\let\auto@bib@innerbib\@empty
\bibitem [{\citenamefont {Zuo}\ \emph {et~al.}(2020)\citenamefont {Zuo},
  \citenamefont {Chen}, \citenamefont {Li}, \citenamefont {Deng}, \citenamefont
  {Chen}, \citenamefont {Behler}, \citenamefont {Cs\'a{}nyi}, \citenamefont
  {Shapeev}, \citenamefont {Thompson}, \citenamefont {Wood},\ and\
  \citenamefont {Ong}}]{Zuo2020}%
  \BibitemOpen
  \bibfield  {author} {\bibinfo {author} {\bibfnamefont {Y.}~\bibnamefont
  {Zuo}}, \bibinfo {author} {\bibfnamefont {C.}~\bibnamefont {Chen}}, \bibinfo
  {author} {\bibfnamefont {X.}~\bibnamefont {Li}}, \bibinfo {author}
  {\bibfnamefont {Z.}~\bibnamefont {Deng}}, \bibinfo {author} {\bibfnamefont
  {Y.}~\bibnamefont {Chen}}, \bibinfo {author} {\bibfnamefont {J.}~\bibnamefont
  {Behler}}, \bibinfo {author} {\bibfnamefont {G.}~\bibnamefont {Cs\'a{}nyi}},
  \bibinfo {author} {\bibfnamefont {A.~V.}\ \bibnamefont {Shapeev}}, \bibinfo
  {author} {\bibfnamefont {A.~P.}\ \bibnamefont {Thompson}}, \bibinfo {author}
  {\bibfnamefont {M.~A.}\ \bibnamefont {Wood}}, \ and\ \bibinfo {author}
  {\bibfnamefont {S.~P.}\ \bibnamefont {Ong}},\ }\href 
  {\doibase 10.1021/acs.jpca.9b08723} {\bibfield
  {journal} {\bibinfo  {journal} {J. Phys. Chem. A}\ }\textbf {\bibinfo
  {volume} {124}},\ \bibinfo {pages} {731} (\bibinfo {year}
  {2020})}\BibitemShut {NoStop}%
\bibitem [{\citenamefont {Behler}\ and\ \citenamefont
  {Parrinello}(2007)}]{Behler2007}%
  \BibitemOpen
  \bibfield  {author} {\bibinfo {author} {\bibfnamefont {J.}~\bibnamefont
  {Behler}}\ and\ \bibinfo {author} {\bibfnamefont {M.}~\bibnamefont
  {Parrinello}},\ }\href {\doibase 10.1103/PhysRevLett.98.146401} {\bibfield
  {journal} {\bibinfo  {journal} {Phys. Rev. Lett.}\ }\textbf {\bibinfo
  {volume} {98}},\ \bibinfo {pages} {146401} (\bibinfo {year}
  {2007})}\BibitemShut {NoStop}%
\bibitem [{\citenamefont {Bart\'ok}\ \emph {et~al.}(2010)\citenamefont
  {Bart\'ok}, \citenamefont {Payne}, \citenamefont {Kondor},\ and\
  \citenamefont {Cs\'anyi}}]{Bartok2010}%
  \BibitemOpen
  \bibfield  {author} {\bibinfo {author} {\bibfnamefont {A.~P.}\ \bibnamefont
  {Bart\'ok}}, \bibinfo {author} {\bibfnamefont {M.~C.}\ \bibnamefont {Payne}},
  \bibinfo {author} {\bibfnamefont {R.}~\bibnamefont {Kondor}}, \ and\ \bibinfo
  {author} {\bibfnamefont {G.}~\bibnamefont {Cs\'anyi}},\ }\href {\doibase
  10.1103/PhysRevLett.104.136403} {\bibfield  {journal} {\bibinfo  {journal}
  {Phys. Rev. Lett.}\ }\textbf {\bibinfo {volume} {104}},\ \bibinfo {pages}
  {136403} (\bibinfo {year} {2010})}\BibitemShut {NoStop}%
\bibitem [{\citenamefont {Thompson}\ \emph {et~al.}(2015)\citenamefont
  {Thompson}, \citenamefont {Swiler}, \citenamefont {Trott}, \citenamefont
  {Foiles},\ and\ \citenamefont {Tucker}}]{Thompson2015}%
  \BibitemOpen
  \bibfield  {author} {\bibinfo {author} {\bibfnamefont {A.}~\bibnamefont
  {Thompson}}, \bibinfo {author} {\bibfnamefont {L.}~\bibnamefont {Swiler}},
  \bibinfo {author} {\bibfnamefont {C.}~\bibnamefont {Trott}}, \bibinfo
  {author} {\bibfnamefont {S.}~\bibnamefont {Foiles}}, \ and\ \bibinfo {author}
  {\bibfnamefont {G.}~\bibnamefont {Tucker}},\ }\href {\doibase
  10.1016/j.jcp.2014.12.018} {\bibfield  {journal} {\bibinfo  {journal} {J.
  Comput. Phys.}\ }\textbf {\bibinfo {volume} {285}},\ \bibinfo {pages} {316}
  (\bibinfo {year} {2015})}\BibitemShut {NoStop}%
\bibitem [{\citenamefont {Li}\ \emph {et~al.}(2015)\citenamefont {Li},
  \citenamefont {Kermode},\ and\ \citenamefont {De~Vita}}]{Li2015}%
  \BibitemOpen
  \bibfield  {author} {\bibinfo {author} {\bibfnamefont {Z.}~\bibnamefont
  {Li}}, \bibinfo {author} {\bibfnamefont {J.~R.}\ \bibnamefont {Kermode}}, \
  and\ \bibinfo {author} {\bibfnamefont {A.}~\bibnamefont {De~Vita}},\ }\href
  {\doibase 10.1103/PhysRevLett.114.096405} {\bibfield  {journal} {\bibinfo
  {journal} {Phys. Rev. Lett.}\ }\textbf {\bibinfo {volume} {114}},\ \bibinfo
  {pages} {096405} (\bibinfo {year} {2015})}\BibitemShut {NoStop}%
\bibitem [{\citenamefont {Shapeev}(2016)}]{Shapeev2016}%
  \BibitemOpen
  \bibfield  {author} {\bibinfo {author} {\bibfnamefont {A.~V.}\ \bibnamefont
  {Shapeev}},\ }\href {\doibase 10.1137/15M1054183} {\bibfield  {journal}
  {\bibinfo  {journal} {Multiscale Model. Simul.}\ }\textbf {\bibinfo {volume}
  {14}},\ \bibinfo {pages} {1153} (\bibinfo {year} {2016})}\BibitemShut
  {NoStop}%
\bibitem [{\citenamefont {Zhang}\ \emph {et~al.}(2018)\citenamefont {Zhang},
  \citenamefont {Han}, \citenamefont {Wang}, \citenamefont {Car},\ and\
  \citenamefont {E}}]{Zhang2018}%
  \BibitemOpen
  \bibfield  {author} {\bibinfo {author} {\bibfnamefont {L.}~\bibnamefont
  {Zhang}}, \bibinfo {author} {\bibfnamefont {J.}~\bibnamefont {Han}}, \bibinfo
  {author} {\bibfnamefont {H.}~\bibnamefont {Wang}}, \bibinfo {author}
  {\bibfnamefont {R.}~\bibnamefont {Car}}, \ and\ \bibinfo {author}
  {\bibfnamefont {W.}~\bibnamefont {E}},\ }\href {\doibase
  10.1103/PhysRevLett.120.143001} {\bibfield  {journal} {\bibinfo  {journal}
  {Phys. Rev. Lett.}\ }\textbf {\bibinfo {volume} {120}},\ \bibinfo {pages}
  {143001} (\bibinfo {year} {2018})}\BibitemShut {NoStop}%
\bibitem [{\citenamefont {Jinnouchi}\ \emph {et~al.}(2019)\citenamefont
  {Jinnouchi}, \citenamefont {Lahnsteiner}, \citenamefont {Karsai},
  \citenamefont {Kresse},\ and\ \citenamefont {Bokdam}}]{Jinnouchi2019}%
  \BibitemOpen
  \bibfield  {author} {\bibinfo {author} {\bibfnamefont {R.}~\bibnamefont
  {Jinnouchi}}, \bibinfo {author} {\bibfnamefont {J.}~\bibnamefont
  {Lahnsteiner}}, \bibinfo {author} {\bibfnamefont {F.}~\bibnamefont {Karsai}},
  \bibinfo {author} {\bibfnamefont {G.}~\bibnamefont {Kresse}}, \ and\ \bibinfo
  {author} {\bibfnamefont {M.}~\bibnamefont {Bokdam}},\ }\href {\doibase
  10.1103/PhysRevLett.122.225701} {\bibfield  {journal} {\bibinfo  {journal}
  {Phys. Rev. Lett.}\ }\textbf {\bibinfo {volume} {122}},\ \bibinfo {pages}
  {225701} (\bibinfo {year} {2019})}\BibitemShut {NoStop}%
\bibitem [{\citenamefont {Sosso}\ \emph {et~al.}(2013)\citenamefont {Sosso},
  \citenamefont {Miceli}, \citenamefont {Caravati}, \citenamefont {Giberti},
  \citenamefont {Behler},\ and\ \citenamefont {Bernasconi}}]{Sosso2013}%
  \BibitemOpen
  \bibfield  {author} {\bibinfo {author} {\bibfnamefont {G.~C.}\ \bibnamefont
  {Sosso}}, \bibinfo {author} {\bibfnamefont {G.}~\bibnamefont {Miceli}},
  \bibinfo {author} {\bibfnamefont {S.}~\bibnamefont {Caravati}}, \bibinfo
  {author} {\bibfnamefont {F.}~\bibnamefont {Giberti}}, \bibinfo {author}
  {\bibfnamefont {J.}~\bibnamefont {Behler}}, \ and\ \bibinfo {author}
  {\bibfnamefont {M.}~\bibnamefont {Bernasconi}},\ }\href {\doibase
  10.1021/jz402268v} {\bibfield  {journal} {\bibinfo  {journal} {J. Phys. Chem.
  Lett.}\ }\textbf {\bibinfo {volume} {4}},\ \bibinfo {pages} {4241} (\bibinfo
  {year} {2013})}\BibitemShut {NoStop}%
\bibitem [{\citenamefont {Caro}\ \emph {et~al.}(2018)\citenamefont {Caro},
  \citenamefont {Deringer}, \citenamefont {Koskinen}, \citenamefont {Laurila},\
  and\ \citenamefont {Cs\'anyi}}]{Caro2018}%
  \BibitemOpen
  \bibfield  {author} {\bibinfo {author} {\bibfnamefont {M.~A.}\ \bibnamefont
  {Caro}}, \bibinfo {author} {\bibfnamefont {V.~L.}\ \bibnamefont {Deringer}},
  \bibinfo {author} {\bibfnamefont {J.}~\bibnamefont {Koskinen}}, \bibinfo
  {author} {\bibfnamefont {T.}~\bibnamefont {Laurila}}, \ and\ \bibinfo
  {author} {\bibfnamefont {G.}~\bibnamefont {Cs\'anyi}},\ }\href {\doibase
  10.1103/PhysRevLett.120.166101} {\bibfield  {journal} {\bibinfo  {journal}
  {Phys. Rev. Lett.}\ }\textbf {\bibinfo {volume} {120}},\ \bibinfo {pages}
  {166101} (\bibinfo {year} {2018})}\BibitemShut {NoStop}%
\bibitem [{\citenamefont {Deringer}\ \emph
  {et~al.}(2021{\natexlab{a}})\citenamefont {Deringer}, \citenamefont
  {Bernstein}, \citenamefont {Cs\'a{}nyi}, \citenamefont {Ben~Mahmoud},
  \citenamefont {Ceriotti}, \citenamefont {Wilson}, \citenamefont {Drabold},\
  and\ \citenamefont {Elliott}}]{aSi_100k}%
  \BibitemOpen
  \bibfield  {author} {\bibinfo {author} {\bibfnamefont {V.~L.}\ \bibnamefont
  {Deringer}}, \bibinfo {author} {\bibfnamefont {N.}~\bibnamefont {Bernstein}},
  \bibinfo {author} {\bibfnamefont {G.}~\bibnamefont {Cs\'a{}nyi}}, \bibinfo
  {author} {\bibfnamefont {C.}~\bibnamefont {Ben~Mahmoud}}, \bibinfo {author}
  {\bibfnamefont {M.}~\bibnamefont {Ceriotti}}, \bibinfo {author}
  {\bibfnamefont {M.}~\bibnamefont {Wilson}}, \bibinfo {author} {\bibfnamefont
  {D.~A.}\ \bibnamefont {Drabold}}, \ and\ \bibinfo {author} {\bibfnamefont
  {S.~R.}\ \bibnamefont {Elliott}},\ }\href {\doibase
  10.1038/s41586-020-03072-z} {\bibfield  {journal} {\bibinfo  {journal}
  {Nature}\ }\textbf {\bibinfo {volume} {589}},\ \bibinfo {pages} {59}
  (\bibinfo {year} {2021}{\natexlab{a}})}\BibitemShut {NoStop}%
\bibitem [{\citenamefont {Cheng}\ \emph
  {et~al.}(2020{\natexlab{a}})\citenamefont {Cheng}, \citenamefont {Mazzola},
  \citenamefont {Pickard},\ and\ \citenamefont {Ceriotti}}]{Cheng2020}%
  \BibitemOpen
  \bibfield  {author} {\bibinfo {author} {\bibfnamefont {B.}~\bibnamefont
  {Cheng}}, \bibinfo {author} {\bibfnamefont {G.}~\bibnamefont {Mazzola}},
  \bibinfo {author} {\bibfnamefont {C.~J.}\ \bibnamefont {Pickard}}, \ and\
  \bibinfo {author} {\bibfnamefont {M.}~\bibnamefont {Ceriotti}},\ }\href
  {\doibase 10.1038/s41586-020-2677-y} {\bibfield  {journal} {\bibinfo
  {journal} {Nature}\ }\textbf {\bibinfo {volume} {585}},\ \bibinfo {pages}
  {217} (\bibinfo {year} {2020}{\natexlab{a}})}\BibitemShut {NoStop}%
\bibitem [{\citenamefont {Smith}\ \emph {et~al.}(2021)\citenamefont {Smith},
  \citenamefont {Nebgen}, \citenamefont {Mathew}, \citenamefont {Chen},
  \citenamefont {Lubbers}, \citenamefont {Burakovsky}, \citenamefont {Tretiak},
  \citenamefont {Nam}, \citenamefont {Germann}, \citenamefont {Fensin},\ and\
  \citenamefont {Barros}}]{Smith2021}%
  \BibitemOpen
  \bibfield  {author} {\bibinfo {author} {\bibfnamefont {J.~S.}\ \bibnamefont
  {Smith}}, \bibinfo {author} {\bibfnamefont {B.}~\bibnamefont {Nebgen}},
  \bibinfo {author} {\bibfnamefont {N.}~\bibnamefont {Mathew}}, \bibinfo
  {author} {\bibfnamefont {J.}~\bibnamefont {Chen}}, \bibinfo {author}
  {\bibfnamefont {N.}~\bibnamefont {Lubbers}}, \bibinfo {author} {\bibfnamefont
  {L.}~\bibnamefont {Burakovsky}}, \bibinfo {author} {\bibfnamefont
  {S.}~\bibnamefont {Tretiak}}, \bibinfo {author} {\bibfnamefont {H.~A.}\
  \bibnamefont {Nam}}, \bibinfo {author} {\bibfnamefont {T.}~\bibnamefont
  {Germann}}, \bibinfo {author} {\bibfnamefont {S.}~\bibnamefont {Fensin}}, \
  and\ \bibinfo {author} {\bibfnamefont {K.}~\bibnamefont {Barros}},\ }\href
  {\doibase 10.1038/s41467-021-21376-0} {\bibfield  {journal} {\bibinfo
  {journal} {Nat. Commun.}\ }\textbf {\bibinfo {volume} {12}},\ \bibinfo
  {pages} {1257} (\bibinfo {year} {2021})}\BibitemShut {NoStop}%
\bibitem [{\citenamefont {Zong}\ \emph {et~al.}(2021)\citenamefont {Zong},
  \citenamefont {Robinson}, \citenamefont {Hermann}, \citenamefont {Zhao},
  \citenamefont {Scandolo}, \citenamefont {Ding},\ and\ \citenamefont
  {Ackland}}]{Zong2021}%
  \BibitemOpen
  \bibfield  {author} {\bibinfo {author} {\bibfnamefont {H.}~\bibnamefont
  {Zong}}, \bibinfo {author} {\bibfnamefont {V.~N.}\ \bibnamefont {Robinson}},
  \bibinfo {author} {\bibfnamefont {A.}~\bibnamefont {Hermann}}, \bibinfo
  {author} {\bibfnamefont {L.}~\bibnamefont {Zhao}}, \bibinfo {author}
  {\bibfnamefont {S.}~\bibnamefont {Scandolo}}, \bibinfo {author}
  {\bibfnamefont {X.}~\bibnamefont {Ding}}, \ and\ \bibinfo {author}
  {\bibfnamefont {G.~J.}\ \bibnamefont {Ackland}},\ }\href {\doibase
  10.1038/s41567-021-01244-w} {\bibfield  {journal} {\bibinfo  {journal} {Nat.
  Phys.}\ }\textbf {\bibinfo {volume} {17}},\ \bibinfo {pages} {955} (\bibinfo
  {year} {2021})}\BibitemShut {NoStop}%
\bibitem [{\citenamefont {Artrith}\ and\ \citenamefont
  {Kolpak}(2014)}]{Artrith2014}%
  \BibitemOpen
  \bibfield  {author} {\bibinfo {author} {\bibfnamefont {N.}~\bibnamefont
  {Artrith}}\ and\ \bibinfo {author} {\bibfnamefont {A.~M.}\ \bibnamefont
  {Kolpak}},\ }\href {\doibase 10.1021/nl5005674} {\bibfield  {journal}
  {\bibinfo  {journal} {Nano Lett.}\ }\textbf {\bibinfo {volume} {14}},\
  \bibinfo {pages} {2670} (\bibinfo {year} {2014})}\BibitemShut {NoStop}%
\bibitem [{\citenamefont {Timmermann}\ \emph {et~al.}(2020)\citenamefont
  {Timmermann}, \citenamefont {Kraushofer}, \citenamefont {Resch},
  \citenamefont {Li}, \citenamefont {Wang}, \citenamefont {Mao}, \citenamefont
  {Riva}, \citenamefont {Lee}, \citenamefont {Staacke}, \citenamefont {Schmid},
  \citenamefont {Scheurer}, \citenamefont {Parkinson}, \citenamefont
  {Diebold},\ and\ \citenamefont {Reuter}}]{Timmermann2020}%
  \BibitemOpen
  \bibfield  {author} {\bibinfo {author} {\bibfnamefont {J.}~\bibnamefont
  {Timmermann}}, \bibinfo {author} {\bibfnamefont {F.}~\bibnamefont
  {Kraushofer}}, \bibinfo {author} {\bibfnamefont {N.}~\bibnamefont {Resch}},
  \bibinfo {author} {\bibfnamefont {P.}~\bibnamefont {Li}}, \bibinfo {author}
  {\bibfnamefont {Y.}~\bibnamefont {Wang}}, \bibinfo {author} {\bibfnamefont
  {Z.}~\bibnamefont {Mao}}, \bibinfo {author} {\bibfnamefont {M.}~\bibnamefont
  {Riva}}, \bibinfo {author} {\bibfnamefont {Y.}~\bibnamefont {Lee}}, \bibinfo
  {author} {\bibfnamefont {C.}~\bibnamefont {Staacke}}, \bibinfo {author}
  {\bibfnamefont {M.}~\bibnamefont {Schmid}}, \bibinfo {author} {\bibfnamefont
  {C.}~\bibnamefont {Scheurer}}, \bibinfo {author} {\bibfnamefont {G.~S.}\
  \bibnamefont {Parkinson}}, \bibinfo {author} {\bibfnamefont {U.}~\bibnamefont
  {Diebold}}, \ and\ \bibinfo {author} {\bibfnamefont {K.}~\bibnamefont
  {Reuter}},\ }\href {\doibase 10.1103/PhysRevLett.125.206101} {\bibfield
  {journal} {\bibinfo  {journal} {Phys. Rev. Lett.}\ }\textbf {\bibinfo
  {volume} {125}},\ \bibinfo {pages} {206101} (\bibinfo {year}
  {2020})}\BibitemShut {NoStop}%
\bibitem [{\citenamefont {Behler}(2017)}]{Behler2017}%
  \BibitemOpen
  \bibfield  {author} {\bibinfo {author} {\bibfnamefont {J.}~\bibnamefont
  {Behler}},\ }\href {\doibase 10.1002/anie.201703114} 
  {\bibfield  {journal} {\bibinfo  {journal} {Angew.
  Chem. Int. Ed.}\ }\textbf {\bibinfo {volume} {56}},\ \bibinfo {pages} {12828}
  (\bibinfo {year} {2017})}\BibitemShut {NoStop}%
\bibitem [{\citenamefont {Deringer}\ \emph {et~al.}(2019)\citenamefont
  {Deringer}, \citenamefont {Caro},\ and\ \citenamefont
  {Cs\'a{}nyi}}]{MLP_AdvMater}%
  \BibitemOpen
  \bibfield  {author} {\bibinfo {author} {\bibfnamefont {V.~L.}\ \bibnamefont
  {Deringer}}, \bibinfo {author} {\bibfnamefont {M.~A.}\ \bibnamefont {Caro}},
  \ and\ \bibinfo {author} {\bibfnamefont {G.}~\bibnamefont {Cs\'a{}nyi}},\
  }\href {\doibase 10.1002/adma.201902765} {\bibfield  {journal} {\bibinfo
  {journal} {Adv. Mater.}\ }\textbf {\bibinfo {volume} {31}},\ \bibinfo {pages}
  {1902765} (\bibinfo {year} {2019})}\BibitemShut {NoStop}%
\bibitem [{\citenamefont {No\'e{}}\ \emph {et~al.}(2020)\citenamefont
  {No\'e{}}, \citenamefont {Tkatchenko}, \citenamefont {M\"u{}ller},\ and\
  \citenamefont {Clementi}}]{Noe2020}%
  \BibitemOpen
  \bibfield  {author} {\bibinfo {author} {\bibfnamefont {F.}~\bibnamefont
  {No\'e{}}}, \bibinfo {author} {\bibfnamefont {A.}~\bibnamefont {Tkatchenko}},
  \bibinfo {author} {\bibfnamefont {K.-R.}\ \bibnamefont {M\"u{}ller}}, \ and\
  \bibinfo {author} {\bibfnamefont {C.}~\bibnamefont {Clementi}},\ }\href
  {\doibase 10.1146/annurev-physchem-042018-052331} {\bibfield  {journal}
  {\bibinfo  {journal} {Annu. Rev. Phys. Chem.}\ }\textbf {\bibinfo {volume}
  {71}},\ \bibinfo {pages} {361} (\bibinfo {year} {2020})}\BibitemShut
  {NoStop}%
\bibitem [{\citenamefont {Friederich}\ \emph {et~al.}(2021)\citenamefont
  {Friederich}, \citenamefont {H\"a{}se}, \citenamefont {Proppe},\ and\
  \citenamefont {Aspuru-Guzik}}]{Friederich2021}%
  \BibitemOpen
  \bibfield  {author} {\bibinfo {author} {\bibfnamefont {P.}~\bibnamefont
  {Friederich}}, \bibinfo {author} {\bibfnamefont {F.}~\bibnamefont
  {H\"a{}se}}, \bibinfo {author} {\bibfnamefont {J.}~\bibnamefont {Proppe}}, \
  and\ \bibinfo {author} {\bibfnamefont {A.}~\bibnamefont {Aspuru-Guzik}},\
  }\href {\doibase 10.1038/s41563-020-0777-6} {\bibfield  {journal} {\bibinfo
  {journal} {Nat. Mater.}\ }\textbf {\bibinfo {volume} {20}},\ \bibinfo {pages}
  {750} (\bibinfo {year} {2021})}\BibitemShut {NoStop}%
\bibitem [{\citenamefont {George}\ \emph {et~al.}(2020)\citenamefont {George},
  \citenamefont {Hautier}, \citenamefont {Bart\'ok}, \citenamefont {Cs\'anyi},\
  and\ \citenamefont {Deringer}}]{George2020}%
  \BibitemOpen
  \bibfield  {author} {\bibinfo {author} {\bibfnamefont {J.}~\bibnamefont
  {George}}, \bibinfo {author} {\bibfnamefont {G.}~\bibnamefont {Hautier}},
  \bibinfo {author} {\bibfnamefont {A.~P.}\ \bibnamefont {Bart\'ok}}, \bibinfo
  {author} {\bibfnamefont {G.}~\bibnamefont {Cs\'anyi}}, \ and\ \bibinfo
  {author} {\bibfnamefont {V.~L.}\ \bibnamefont {Deringer}},\ }\href {\doibase
  10.1063/5.0013826} {\bibfield  {journal} {\bibinfo  {journal} {J. Chem.
  Phys.}\ }\textbf {\bibinfo {volume} {153}},\ \bibinfo {pages} {044104}
  (\bibinfo {year} {2020})}\BibitemShut {NoStop}%
\bibitem [{\citenamefont {{de Tomas}}\ \emph {et~al.}(2019)\citenamefont {{de
  Tomas}}, \citenamefont {Aghajamali}, \citenamefont {Jones}, \citenamefont
  {Lim}, \citenamefont {L\'o{}pez}, \citenamefont {Suarez-Martinez},\ and\
  \citenamefont {Marks}}]{deTomas2019}%
  \BibitemOpen
  \bibfield  {author} {\bibinfo {author} {\bibfnamefont {C.}~\bibnamefont {{de
  Tomas}}}, \bibinfo {author} {\bibfnamefont {A.}~\bibnamefont {Aghajamali}},
  \bibinfo {author} {\bibfnamefont {J.~L.}\ \bibnamefont {Jones}}, \bibinfo
  {author} {\bibfnamefont {D.~J.}\ \bibnamefont {Lim}}, \bibinfo {author}
  {\bibfnamefont {M.~J.}\ \bibnamefont {L\'o{}pez}}, \bibinfo {author}
  {\bibfnamefont {I.}~\bibnamefont {Suarez-Martinez}}, \ and\ \bibinfo {author}
  {\bibfnamefont {N.~A.}\ \bibnamefont {Marks}},\ }\href {\doibase
  10.1016/j.carbon.2019.07.074} {\bibfield  {journal} {\bibinfo  {journal}
  {Carbon}\ }\textbf {\bibinfo {volume} {155}},\ \bibinfo {pages} {624}
  (\bibinfo {year} {2019})}\BibitemShut {NoStop}%
\bibitem [{\citenamefont {Bart\'ok}\ \emph
  {et~al.}(2018{\natexlab{a}})\citenamefont {Bart\'ok}, \citenamefont
  {Kermode}, \citenamefont {Bernstein},\ and\ \citenamefont
  {Cs\'a{}nyi}}]{Bartok2018}%
  \BibitemOpen
  \bibfield  {author} {\bibinfo {author} {\bibfnamefont {A.~P.}\ \bibnamefont
  {Bart\'ok}}, \bibinfo {author} {\bibfnamefont {J.}~\bibnamefont {Kermode}},
  \bibinfo {author} {\bibfnamefont {N.}~\bibnamefont {Bernstein}}, \ and\
  \bibinfo {author} {\bibfnamefont {G.}~\bibnamefont {Cs\'a{}nyi}},\ }\href
  {\doibase 10.1103/PhysRevX.8.041048} {\bibfield  {journal} {\bibinfo
  {journal} {Phys. Rev. X}\ }\textbf {\bibinfo {volume} {8}},\ \bibinfo {pages}
  {041048} (\bibinfo {year} {2018}{\natexlab{a}})}\BibitemShut {NoStop}%
\bibitem [{\citenamefont {Lysogorskiy}\ \emph {et~al.}(2021)\citenamefont
  {Lysogorskiy}, \citenamefont {van~der Oord}, \citenamefont {Bochkarev},
  \citenamefont {Menon}, \citenamefont {Rinaldi}, \citenamefont
  {Hammerschmidt}, \citenamefont {Mrovec}, \citenamefont {Thompson},
  \citenamefont {Cs\'a{}nyi}, \citenamefont {Ortner},\ and\ \citenamefont
  {Drautz}}]{Lysogorskiy2021}%
  \BibitemOpen
  \bibfield  {author} {\bibinfo {author} {\bibfnamefont {Y.}~\bibnamefont
  {Lysogorskiy}}, \bibinfo {author} {\bibfnamefont {C.}~\bibnamefont {van~der
  Oord}}, \bibinfo {author} {\bibfnamefont {A.}~\bibnamefont {Bochkarev}},
  \bibinfo {author} {\bibfnamefont {S.}~\bibnamefont {Menon}}, \bibinfo
  {author} {\bibfnamefont {M.}~\bibnamefont {Rinaldi}}, \bibinfo {author}
  {\bibfnamefont {T.}~\bibnamefont {Hammerschmidt}}, \bibinfo {author}
  {\bibfnamefont {M.}~\bibnamefont {Mrovec}}, \bibinfo {author} {\bibfnamefont
  {A.}~\bibnamefont {Thompson}}, \bibinfo {author} {\bibfnamefont
  {G.}~\bibnamefont {Cs\'a{}nyi}}, \bibinfo {author} {\bibfnamefont
  {C.}~\bibnamefont {Ortner}}, \ and\ \bibinfo {author} {\bibfnamefont
  {R.}~\bibnamefont {Drautz}},\ }\href {\doibase 10.1038/s41524-021-00559-9}
  {\bibfield  {journal} {\bibinfo  {journal} {npj Comput. Mater.}\ }\textbf
  {\bibinfo {volume} {7}},\ \bibinfo {pages} {97} (\bibinfo {year}
  {2021})}\BibitemShut {NoStop}%
\bibitem [{\citenamefont {Hinton}, \citenamefont {Vinyals},\ and\ \citenamefont
  {Dean}()}]{Hinton2015}%
  \BibitemOpen
  \bibfield  {author} {\bibinfo {author} {\bibfnamefont {G.}~\bibnamefont
  {Hinton}}, \bibinfo {author} {\bibfnamefont {O.}~\bibnamefont {Vinyals}}, \
  and\ \bibinfo {author} {\bibfnamefont {J.}~\bibnamefont {Dean}},\ }\href@noop
  {} {\enquote {\bibinfo {title} {Distilling the knowledge in a neural
  network},}\ }\Eprint {http://arxiv.org/abs/1503.02531} {arXiv:1503.02531
  [stat.ML]} \BibitemShut {NoStop}%
\bibitem [{\citenamefont {Nguyen}, \citenamefont {Chen},\ and\ \citenamefont
  {Lee}()}]{Nguyen2021}%
  \BibitemOpen
  \bibfield  {author} {\bibinfo {author} {\bibfnamefont {T.}~\bibnamefont
  {Nguyen}}, \bibinfo {author} {\bibfnamefont {Z.}~\bibnamefont {Chen}}, \ and\
  \bibinfo {author} {\bibfnamefont {J.}~\bibnamefont {Lee}},\ }\href
  {https://openreview.net/forum?id=l-PrrQrK0QR} {\ }\bibinfo {note} {``Dataset
  Meta-Learning from Kernel Ridge-Regression" (International Conference on
  Learning Representations, Vienna, 2021)}\BibitemShut {NoStop}%
\bibitem{LopezdePrado2018}%
  \BibitemOpen
  \bibfield  {author} {\bibinfo {author} {\bibfnamefont {M.}~\bibnamefont
  {L\'o{}pez de Prado}},\ }
  \bibinfo {note} { {\em Advances in financial machine learning} 
  (John Wiley \& Sons, Hoboken, NJ, 2018)}\BibitemShut {NoStop}%
\bibitem [{\citenamefont {Vilalta}\ and\ \citenamefont
  {Drissi}(2002)}]{Vilalta2002}%
  \BibitemOpen
  \bibfield  {author} {\bibinfo {author} {\bibfnamefont {R.}~\bibnamefont
  {Vilalta}}\ and\ \bibinfo {author} {\bibfnamefont {Y.}~\bibnamefont
  {Drissi}},\ }\href {\doibase 10.1023/A:1019956318069} 
  {\bibfield  {journal} {\bibinfo  {journal} {Artif.
  Intell. Rev.}\ }\textbf {\bibinfo {volume} {18}},\ \bibinfo {pages} {77}
  (\bibinfo {year} {2002})}\BibitemShut {NoStop}%
\bibitem [{\citenamefont {Freund}\ and\ \citenamefont
  {Schapire}(1997)}]{Freund1997}%
  \BibitemOpen
  \bibfield  {author} {\bibinfo {author} {\bibfnamefont {Y.}~\bibnamefont
  {Freund}}\ and\ \bibinfo {author} {\bibfnamefont {R.~E.}\ \bibnamefont
  {Schapire}},\ }\href {\doibase 10.1006/JCSS.1997.1504} {\bibfield  {journal}
  {\bibinfo  {journal} {J. Comput. Syst. Sci.}\ }\textbf {\bibinfo {volume}
  {55}},\ \bibinfo {pages} {119} (\bibinfo {year} {1997})}\BibitemShut
  {NoStop}%
\bibitem [{\citenamefont {Friedman}\ \emph {et~al.}(2000)\citenamefont
  {Friedman}, \citenamefont {Hastie},\ and\ \citenamefont
  {Tibshirani}}]{Friedman2000}%
  \BibitemOpen
  \bibfield  {author} {\bibinfo {author} {\bibfnamefont {J.}~\bibnamefont
  {Friedman}}, \bibinfo {author} {\bibfnamefont {T.}~\bibnamefont {Hastie}}, \
  and\ \bibinfo {author} {\bibfnamefont {R.}~\bibnamefont {Tibshirani}},\
  }\href {\doibase 10.1214/aos/1016218223} {\bibfield  {journal} {\bibinfo
  {journal} {Ann. Stat.}\ }\textbf {\bibinfo {volume} {28}},\ \bibinfo {pages}
  {337} (\bibinfo {year} {2000})}\BibitemShut {NoStop}%
\bibitem [{\citenamefont {Dietterich}(2000)}]{Dietterich2000}%
  \BibitemOpen
  \bibfield  {author} {\bibinfo {author} {\bibfnamefont {T.~G.}\ \bibnamefont
  {Dietterich}},\ }in\ \href@noop {} {\emph {\bibinfo {booktitle} {Multiple
  Classifier Systems}}}\ (\bibinfo  {publisher} {Springer},\ \bibinfo {address}
  {Berlin, Heidelberg},\ \bibinfo {year} {2000})\ pp.\ \bibinfo {pages}
  {1--15}\BibitemShut {NoStop}%
\bibitem [{\citenamefont {Deringer}\ \emph
  {et~al.}(2021{\natexlab{b}})\citenamefont {Deringer}, \citenamefont
  {Bart\'o{}k}, \citenamefont {Bernstein}, \citenamefont {Wilkins},
  \citenamefont {Ceriotti},\ and\ \citenamefont {Cs\'anyi}}]{chemrev}%
  \BibitemOpen
  \bibfield  {author} {\bibinfo {author} {\bibfnamefont {V.~L.}\ \bibnamefont
  {Deringer}}, \bibinfo {author} {\bibfnamefont {A.~P.}\ \bibnamefont
  {Bart\'o{}k}}, \bibinfo {author} {\bibfnamefont {N.}~\bibnamefont
  {Bernstein}}, \bibinfo {author} {\bibfnamefont {D.~M.}\ \bibnamefont
  {Wilkins}}, \bibinfo {author} {\bibfnamefont {M.}~\bibnamefont {Ceriotti}}, \
  and\ \bibinfo {author} {\bibfnamefont {G.}~\bibnamefont {Cs\'anyi}},\ }\href
  {\doibase 10.1021/acs.chemrev.1c00022} {\bibfield  {journal} {\bibinfo
  {journal} {Chem. Rev.}\ }\textbf {\bibinfo {volume} {121}},\ \bibinfo {pages}
  {10073} (\bibinfo {year} {2021}{\natexlab{b}})}\BibitemShut {NoStop}%
\bibitem [{\citenamefont {Musil}\ \emph {et~al.}(2021)\citenamefont {Musil},
  \citenamefont {Grisafi}, \citenamefont {Bart\'o{}k}, \citenamefont {Ortner},
  \citenamefont {Cs\'a{}nyi},\ and\ \citenamefont {Ceriotti}}]{Musil2021}%
  \BibitemOpen
  \bibfield  {author} {\bibinfo {author} {\bibfnamefont {F.}~\bibnamefont
  {Musil}}, \bibinfo {author} {\bibfnamefont {A.}~\bibnamefont {Grisafi}},
  \bibinfo {author} {\bibfnamefont {A.~P.}\ \bibnamefont {Bart\'o{}k}},
  \bibinfo {author} {\bibfnamefont {C.}~\bibnamefont {Ortner}}, \bibinfo
  {author} {\bibfnamefont {G.}~\bibnamefont {Cs\'a{}nyi}}, \ and\ \bibinfo
  {author} {\bibfnamefont {M.}~\bibnamefont {Ceriotti}},\ }\href {\doibase
  10.1021/acs.chemrev.1c00021} {\bibfield  {journal} {\bibinfo  {journal}
  {Chem. Rev.}\ }\textbf {\bibinfo {volume} {121}},\ \bibinfo {pages} {9759}
  (\bibinfo {year} {2021})}\BibitemShut {NoStop}%
\bibitem [{Note1()}]{Note1}%
  \BibitemOpen
  \bibinfo {note} {In practice, the matter is more complex: typically, atomic
  energies are machine-learned from total energies (and, often, forces) \cite
  {Behler2007, Bartok2010}.}\BibitemShut {Stop}%
\bibitem [{\citenamefont {Cheng}\ \emph
  {et~al.}(2020{\natexlab{b}})\citenamefont {Cheng}, \citenamefont {Griffiths},
  \citenamefont {Wengert}, \citenamefont {Kunkel}, \citenamefont {Stenczel},
  \citenamefont {Zhu}, \citenamefont {Deringer}, \citenamefont {Bernstein},
  \citenamefont {Margraf}, \citenamefont {Reuter},\ and\ \citenamefont
  {Csanyi}}]{Cheng2020a}%
  \BibitemOpen
  \bibfield  {author} {\bibinfo {author} {\bibfnamefont {B.}~\bibnamefont
  {Cheng}}, \bibinfo {author} {\bibfnamefont {R.-R.}\ \bibnamefont
  {Griffiths}}, \bibinfo {author} {\bibfnamefont {S.}~\bibnamefont {Wengert}},
  \bibinfo {author} {\bibfnamefont {C.}~\bibnamefont {Kunkel}}, \bibinfo
  {author} {\bibfnamefont {T.}~\bibnamefont {Stenczel}}, \bibinfo {author}
  {\bibfnamefont {B.}~\bibnamefont {Zhu}}, \bibinfo {author} {\bibfnamefont
  {V.~L.}\ \bibnamefont {Deringer}}, \bibinfo {author} {\bibfnamefont
  {N.}~\bibnamefont {Bernstein}}, \bibinfo {author} {\bibfnamefont {J.~T.}\
  \bibnamefont {Margraf}}, \bibinfo {author} {\bibfnamefont {K.}~\bibnamefont
  {Reuter}}, \ and\ \bibinfo {author} {\bibfnamefont {G.}~\bibnamefont
  {Csanyi}},\ }\href {\doibase 10.1021/acs.accounts.0c00403} {\bibfield
  {journal} {\bibinfo  {journal} {Acc. Chem. Res.}\ }\textbf {\bibinfo {volume}
  {53}},\ \bibinfo {pages} {1981} (\bibinfo {year}
  {2020}{\natexlab{b}})}\BibitemShut {NoStop}%
\bibitem [{\citenamefont {Bart\'ok}\ \emph
  {et~al.}(2018{\natexlab{b}})\citenamefont {Bart\'ok}, \citenamefont
  {Kermode}, \citenamefont {Bernstein},\ and\ \citenamefont
  {Cs\'anyi}}]{GAP-18}%
  \BibitemOpen
  \bibfield  {author} {\bibinfo {author} {\bibfnamefont {A.~P.}\ \bibnamefont
  {Bart\'ok}}, \bibinfo {author} {\bibfnamefont {J.}~\bibnamefont {Kermode}},
  \bibinfo {author} {\bibfnamefont {N.}~\bibnamefont {Bernstein}}, \ and\
  \bibinfo {author} {\bibfnamefont {G.}~\bibnamefont {Cs\'anyi}},\ }\href
  {\doibase 10.1103/PhysRevX.8.041048} {\bibfield  {journal} {\bibinfo
  {journal} {Phys. Rev. X}\ }\textbf {\bibinfo {volume} {8}},\ \bibinfo {pages}
  {041048} (\bibinfo {year} {2018}{\natexlab{b}})}\BibitemShut {NoStop}%
\bibitem [{\citenamefont {Bart\'ok}\ \emph {et~al.}(2013)\citenamefont
  {Bart\'ok}, \citenamefont {Kondor},\ and\ \citenamefont
  {Cs\'anyi}}]{Bartok2013}%
  \BibitemOpen
  \bibfield  {author} {\bibinfo {author} {\bibfnamefont {A.~P.}\ \bibnamefont
  {Bart\'ok}}, \bibinfo {author} {\bibfnamefont {R.}~\bibnamefont {Kondor}}, \
  and\ \bibinfo {author} {\bibfnamefont {G.}~\bibnamefont {Cs\'anyi}},\ }\href
  {\doibase 10.1103/PhysRevB.87.184115} {\bibfield  {journal} {\bibinfo
  {journal} {Phys. Rev. B}\ }\textbf {\bibinfo {volume} {87}},\ \bibinfo
  {pages} {184115} (\bibinfo {year} {2013})}\BibitemShut {NoStop}%
\bibitem [{\citenamefont {Novikov}\ \emph {et~al.}(2021)\citenamefont
  {Novikov}, \citenamefont {Gubaev}, \citenamefont {Podryabinkin},\ and\
  \citenamefont {Shapeev}}]{Novikov2021}%
  \BibitemOpen
  \bibfield  {author} {\bibinfo {author} {\bibfnamefont {I.~S.}\ \bibnamefont
  {Novikov}}, \bibinfo {author} {\bibfnamefont {K.}~\bibnamefont {Gubaev}},
  \bibinfo {author} {\bibfnamefont {E.~V.}\ \bibnamefont {Podryabinkin}}, \
  and\ \bibinfo {author} {\bibfnamefont {A.~V.}\ \bibnamefont {Shapeev}},\
  }\href {\doibase 10.1088/2632-2153/abc9fe} {\bibfield  {journal} {\bibinfo
  {journal} {Mach. Learn.: Sci. Technol.}\ }\textbf {\bibinfo {volume} {2}},\
  \bibinfo {pages} {025002} (\bibinfo {year} {2021})}\BibitemShut {NoStop}%
\bibitem [{Note2()}]{Note2}%
  \BibitemOpen
  \bibinfo {note} {We note that the idea of mapping a kernel-based ML potential
  to a fast approximator has been previously explored: via cubic spline
  interpolation across a pre-calculated grid of points \cite{Glielmo2018,
  Vandermause2020}.}\BibitemShut {Stop}%
\bibitem [{\citenamefont {Pickard}\ and\ \citenamefont
  {Needs}(2006)}]{Pickard2006}%
  \BibitemOpen
  \bibfield  {author} {\bibinfo {author} {\bibfnamefont {C.~J.}\ \bibnamefont
  {Pickard}}\ and\ \bibinfo {author} {\bibfnamefont {R.~J.}\ \bibnamefont
  {Needs}},\ }\href {\doibase 10.1103/PhysRevLett.97.045504} {\bibfield
  {journal} {\bibinfo  {journal} {Phys. Rev. Lett.}\ }\textbf {\bibinfo
  {volume} {97}},\ \bibinfo {pages} {045504} (\bibinfo {year}
  {2006})}\BibitemShut {NoStop}%
\bibitem [{\citenamefont {Pickard}\ and\ \citenamefont
  {Needs}(2011)}]{Pickard2011}%
  \BibitemOpen
  \bibfield  {author} {\bibinfo {author} {\bibfnamefont {C.~J.}\ \bibnamefont
  {Pickard}}\ and\ \bibinfo {author} {\bibfnamefont {R.~J.}\ \bibnamefont
  {Needs}},\ }\href {\doibase 10.1088/0953-8984/23/5/053201} {\bibfield
  {journal} {\bibinfo  {journal} {J. Phys.: Condens. Matter}\ }\textbf
  {\bibinfo {volume} {23}},\ \bibinfo {pages} {53201} (\bibinfo {year}
  {2011})}\BibitemShut {NoStop}%
\bibitem [{\citenamefont {Perdew}\ \emph {et~al.}(1992)\citenamefont {Perdew},
  \citenamefont {Chevary}, \citenamefont {Vosko}, \citenamefont {Jackson},
  \citenamefont {Pederson}, \citenamefont {Singh},\ and\ \citenamefont
  {Fiolhais}}]{Perdew1992}%
  \BibitemOpen
  \bibfield  {author} {\bibinfo {author} {\bibfnamefont {J.~P.}\ \bibnamefont
  {Perdew}}, \bibinfo {author} {\bibfnamefont {J.~A.}\ \bibnamefont {Chevary}},
  \bibinfo {author} {\bibfnamefont {S.~H.}\ \bibnamefont {Vosko}}, \bibinfo
  {author} {\bibfnamefont {K.~A.}\ \bibnamefont {Jackson}}, \bibinfo {author}
  {\bibfnamefont {M.~R.}\ \bibnamefont {Pederson}}, \bibinfo {author}
  {\bibfnamefont {D.~J.}\ \bibnamefont {Singh}}, \ and\ \bibinfo {author}
  {\bibfnamefont {C.}~\bibnamefont {Fiolhais}},\ }\href {\doibase
  10.1103/PhysRevB.46.6671} {\bibfield  {journal} {\bibinfo  {journal} {Phys.
  Rev. B}\ }\textbf {\bibinfo {volume} {46}},\ \bibinfo {pages} {6671}
  (\bibinfo {year} {1992})}\BibitemShut {NoStop}%
\bibitem [{\citenamefont {Clark}\ \emph {et~al.}(2005)\citenamefont {Clark},
  \citenamefont {Segall}, \citenamefont {Pickard}, \citenamefont {Hasnip},
  \citenamefont {Probert}, \citenamefont {Refson},\ and\ \citenamefont
  {Payne}}]{Clark2005}%
  \BibitemOpen
  \bibfield  {author} {\bibinfo {author} {\bibfnamefont {S.~J.}\ \bibnamefont
  {Clark}}, \bibinfo {author} {\bibfnamefont {M.~D.}\ \bibnamefont {Segall}},
  \bibinfo {author} {\bibfnamefont {C.~J.}\ \bibnamefont {Pickard}}, \bibinfo
  {author} {\bibfnamefont {P.~J.}\ \bibnamefont {Hasnip}}, \bibinfo {author}
  {\bibfnamefont {M.~I.~J.}\ \bibnamefont {Probert}}, \bibinfo {author}
  {\bibfnamefont {K.}~\bibnamefont {Refson}}, \ and\ \bibinfo {author}
  {\bibfnamefont {M.~C.}\ \bibnamefont {Payne}},\ }\href {\doibase
  doi:10.1524/zkri.220.5.567.65075} {\bibfield  {journal} {\bibinfo  {journal}
  {Z. Kristallogr.}\ }\textbf {\bibinfo {volume} {220}},\ \bibinfo {pages}
  {567} (\bibinfo {year} {2005})}\BibitemShut {NoStop}%
\bibitem [{\citenamefont {Rosenbrock}\ \emph {et~al.}(2021)\citenamefont
  {Rosenbrock}, \citenamefont {Gubaev}, \citenamefont {Shapeev}, \citenamefont
  {P{\'{a}}rtay}, \citenamefont {Bernstein}, \citenamefont {Cs{\'{a}}nyi},\
  and\ \citenamefont {Hart}}]{Rosenbrock2021}%
  \BibitemOpen
  \bibfield  {author} {\bibinfo {author} {\bibfnamefont {C.~W.}\ \bibnamefont
  {Rosenbrock}}, \bibinfo {author} {\bibfnamefont {K.}~\bibnamefont {Gubaev}},
  \bibinfo {author} {\bibfnamefont {A.~V.}\ \bibnamefont {Shapeev}}, \bibinfo
  {author} {\bibfnamefont {L.~B.}\ \bibnamefont {P{\'{a}}rtay}}, \bibinfo
  {author} {\bibfnamefont {N.}~\bibnamefont {Bernstein}}, \bibinfo {author}
  {\bibfnamefont {G.}~\bibnamefont {Cs{\'{a}}nyi}}, \ and\ \bibinfo {author}
  {\bibfnamefont {G.~L.~W.}\ \bibnamefont {Hart}},\ }\href {\doibase
  10.1038/s41524-020-00477-2} {\bibfield  {journal} {\bibinfo  {journal} {npj
  Comput. Mater.}\ }\textbf {\bibinfo {volume} {7}},\ \bibinfo {pages} {1}
  (\bibinfo {year} {2021})}\BibitemShut {NoStop}%
\bibitem [{\citenamefont {Deb}\ \emph {et~al.}(2001)\citenamefont {Deb},
  \citenamefont {Wilding}, \citenamefont {Somayazulu},\ and\ \citenamefont
  {McMillan}}]{Deb2001}%
  \BibitemOpen
  \bibfield  {author} {\bibinfo {author} {\bibfnamefont {S.~K.}\ \bibnamefont
  {Deb}}, \bibinfo {author} {\bibfnamefont {M.}~\bibnamefont {Wilding}},
  \bibinfo {author} {\bibfnamefont {M.}~\bibnamefont {Somayazulu}}, \ and\
  \bibinfo {author} {\bibfnamefont {P.~F.}\ \bibnamefont {McMillan}},\ }\href
  {\doibase 10.1038/35107036} {\bibfield  {journal} {\bibinfo  {journal}
  {Nature}\ }\textbf {\bibinfo {volume} {414}},\ \bibinfo {pages} {528}
  (\bibinfo {year} {2001})}\BibitemShut {NoStop}%
\bibitem [{\citenamefont {Wilding}\ \emph {et~al.}(2006)\citenamefont
  {Wilding}, \citenamefont {Wilson},\ and\ \citenamefont
  {McMillan}}]{Wilding2006}%
  \BibitemOpen
  \bibfield  {author} {\bibinfo {author} {\bibfnamefont {M.~C.}\ \bibnamefont
  {Wilding}}, \bibinfo {author} {\bibfnamefont {M.}~\bibnamefont {Wilson}}, \
  and\ \bibinfo {author} {\bibfnamefont {P.~F.}\ \bibnamefont {McMillan}},\
  }\href {\doibase 10.1039/B517775H} {\bibfield  {journal} {\bibinfo  {journal}
  {Chem. Soc. Rev.}\ }\textbf {\bibinfo {volume} {35}},\ \bibinfo {pages} {964}
  (\bibinfo {year} {2006})}\BibitemShut {NoStop}%
\bibitem [{\citenamefont {McMillan}(2021)}]{McMillan2021}%
  \BibitemOpen
  \bibfield  {author} {\bibinfo {author} {\bibfnamefont {P.~F.}\ \bibnamefont
  {McMillan}},\ }\href {\doibase 10.1038/d41586-020-03574-w} {\bibfield
  {journal} {\bibinfo  {journal} {Nature}\ }\textbf {\bibinfo {volume} {589}},\
  \bibinfo {pages} {22} (\bibinfo {year} {2021})}\BibitemShut {NoStop}%
\bibitem [{\citenamefont {Podryabinkin}\ and\ \citenamefont
  {Shapeev}(2017)}]{Podryabinkin2017}%
  \BibitemOpen
  \bibfield  {author} {\bibinfo {author} {\bibfnamefont {E.~V.}\ \bibnamefont
  {Podryabinkin}}\ and\ \bibinfo {author} {\bibfnamefont {A.~V.}\ \bibnamefont
  {Shapeev}},\ }\href {\doibase
  https://doi.org/10.1016/j.commatsci.2017.08.031} {\bibfield  {journal}
  {\bibinfo  {journal} {Comput. Mater. Sci.}\ }\textbf {\bibinfo {volume}
  {140}},\ \bibinfo {pages} {171} (\bibinfo {year} {2017})}\BibitemShut
  {NoStop}%
\bibitem [{\citenamefont {Podryabinkin}\ \emph {et~al.}(2019)\citenamefont
  {Podryabinkin}, \citenamefont {Tikhonov}, \citenamefont {Shapeev},\ and\
  \citenamefont {Oganov}}]{Podryabinkin2019}%
  \BibitemOpen
  \bibfield  {author} {\bibinfo {author} {\bibfnamefont {E.~V.}\ \bibnamefont
  {Podryabinkin}}, \bibinfo {author} {\bibfnamefont {E.~V.}\ \bibnamefont
  {Tikhonov}}, \bibinfo {author} {\bibfnamefont {A.~V.}\ \bibnamefont
  {Shapeev}}, \ and\ \bibinfo {author} {\bibfnamefont {A.~R.}\ \bibnamefont
  {Oganov}},\ }\href {\doibase 10.1103/PhysRevB.99.064114} {\bibfield
  {journal} {\bibinfo  {journal} {Phys. Rev. B}\ }\textbf {\bibinfo {volume}
  {99}},\ \bibinfo {pages} {064114} (\bibinfo {year} {2019})}\BibitemShut
  {NoStop}%
\bibitem [{\citenamefont {Bernstein}\ \emph {et~al.}(2019)\citenamefont
  {Bernstein}, \citenamefont {Bhattarai}, \citenamefont {Cs\'a{}nyi},
  \citenamefont {Drabold}, \citenamefont {Elliott},\ and\ \citenamefont
  {Deringer}}]{Bernstein2019a}%
  \BibitemOpen
  \bibfield  {author} {\bibinfo {author} {\bibfnamefont {N.}~\bibnamefont
  {Bernstein}}, \bibinfo {author} {\bibfnamefont {B.}~\bibnamefont
  {Bhattarai}}, \bibinfo {author} {\bibfnamefont {G.}~\bibnamefont
  {Cs\'a{}nyi}}, \bibinfo {author} {\bibfnamefont {D.~A.}\ \bibnamefont
  {Drabold}}, \bibinfo {author} {\bibfnamefont {S.~R.}\ \bibnamefont
  {Elliott}}, \ and\ \bibinfo {author} {\bibfnamefont {V.~L.}\ \bibnamefont
  {Deringer}},\ }\href {\doibase https://doi.org/10.1002/anie.201902625}
  {\bibfield  {journal} {\bibinfo  {journal} {Angew. Chem. Int. Ed.}\ }\textbf
  {\bibinfo {volume} {58}},\ \bibinfo {pages} {7057} (\bibinfo {year}
  {2019})}\BibitemShut {NoStop}%
\bibitem [{\citenamefont {Xie}\ \emph {et~al.}(2013)\citenamefont {Xie},
  \citenamefont {Long}, \citenamefont {Weigand}, \citenamefont {Moss},
  \citenamefont {Carvalho}, \citenamefont {Roorda}, \citenamefont {Hejna},
  \citenamefont {Torquato},\ and\ \citenamefont {Steinhardt}}]{Xie2013}%
  \BibitemOpen
  \bibfield  {author} {\bibinfo {author} {\bibfnamefont {R.}~\bibnamefont
  {Xie}}, \bibinfo {author} {\bibfnamefont {G.~G.}\ \bibnamefont {Long}},
  \bibinfo {author} {\bibfnamefont {S.~J.}\ \bibnamefont {Weigand}}, \bibinfo
  {author} {\bibfnamefont {S.~C.}\ \bibnamefont {Moss}}, \bibinfo {author}
  {\bibfnamefont {T.}~\bibnamefont {Carvalho}}, \bibinfo {author}
  {\bibfnamefont {S.}~\bibnamefont {Roorda}}, \bibinfo {author} {\bibfnamefont
  {M.}~\bibnamefont {Hejna}}, \bibinfo {author} {\bibfnamefont
  {S.}~\bibnamefont {Torquato}}, \ and\ \bibinfo {author} {\bibfnamefont
  {P.~J.}\ \bibnamefont {Steinhardt}},\ }\href {\doibase
  10.1073/pnas.1220106110} {\bibfield  {journal} {\bibinfo  {journal} {Proc.
  Natl. Acad. Sci. U. S. A.}\ }\textbf {\bibinfo {volume} {110}},\ \bibinfo
  {pages} {13250} (\bibinfo {year} {2013})}\BibitemShut {NoStop}%
\bibitem [{\citenamefont {Plimpton}(1995)}]{Plimpton1995}%
  \BibitemOpen
  \bibfield  {author} {\bibinfo {author} {\bibfnamefont {S.}~\bibnamefont
  {Plimpton}},\ }\href {\doibase 10.1006/jcph.1995.1039} {\bibfield  {journal}
  {\bibinfo  {journal} {J. Comput. Phys.}\ }\textbf {\bibinfo {volume} {117}},\
  \bibinfo {pages} {1} (\bibinfo {year} {1995})}\BibitemShut {NoStop}%
\bibitem [{\citenamefont {Atta-Fynn}\ and\ \citenamefont
  {Biswas}(2018)}]{AttaFynn2018}%
  \BibitemOpen
  \bibfield  {author} {\bibinfo {author} {\bibfnamefont {R.}~\bibnamefont
  {Atta-Fynn}}\ and\ \bibinfo {author} {\bibfnamefont {P.}~\bibnamefont
  {Biswas}},\ }\href {\doibase 10.1063/1.5021813} {\bibfield  {journal}
  {\bibinfo  {journal} {J. Chem. Phys.}\ }\textbf {\bibinfo {volume} {148}},\
  \bibinfo {pages} {204503} (\bibinfo {year} {2018})}\BibitemShut {NoStop}%
\bibitem [{\citenamefont {Deringer}\ \emph {et~al.}(2018)\citenamefont
  {Deringer}, \citenamefont {Bernstein}, \citenamefont {Bart\'o{}k},
  \citenamefont {Cliffe}, \citenamefont {Kerber}, \citenamefont {Marbella},
  \citenamefont {Grey}, \citenamefont {Elliott},\ and\ \citenamefont
  {Cs\'a{}nyi}}]{aSi_structures_JPCL}%
  \BibitemOpen
  \bibfield  {author} {\bibinfo {author} {\bibfnamefont {V.~L.}\ \bibnamefont
  {Deringer}}, \bibinfo {author} {\bibfnamefont {N.}~\bibnamefont {Bernstein}},
  \bibinfo {author} {\bibfnamefont {A.~P.}\ \bibnamefont {Bart\'o{}k}},
  \bibinfo {author} {\bibfnamefont {M.~J.}\ \bibnamefont {Cliffe}}, \bibinfo
  {author} {\bibfnamefont {R.~N.}\ \bibnamefont {Kerber}}, \bibinfo {author}
  {\bibfnamefont {L.~E.}\ \bibnamefont {Marbella}}, \bibinfo {author}
  {\bibfnamefont {C.~P.}\ \bibnamefont {Grey}}, \bibinfo {author}
  {\bibfnamefont {S.~R.}\ \bibnamefont {Elliott}}, \ and\ \bibinfo {author}
  {\bibfnamefont {G.}~\bibnamefont {Cs\'a{}nyi}},\ }\href {\doibase
  10.1021/acs.jpclett.8b00902} {\bibfield  {journal} {\bibinfo  {journal} {J.
  Phys. Chem. Lett.}\ }\textbf {\bibinfo {volume} {9}},\ \bibinfo {pages}
  {2879} (\bibinfo {year} {2018})}\BibitemShut {NoStop}%
\bibitem [{\citenamefont {Limbu}\ \emph {et~al.}(2020)\citenamefont {Limbu},
  \citenamefont {Elliott}, \citenamefont {Atta-Fynn},\ and\ \citenamefont
  {Biswas}}]{Limbu2020}%
  \BibitemOpen
  \bibfield  {author} {\bibinfo {author} {\bibfnamefont {D.~K.}\ \bibnamefont
  {Limbu}}, \bibinfo {author} {\bibfnamefont {S.~R.}\ \bibnamefont {Elliott}},
  \bibinfo {author} {\bibfnamefont {R.}~\bibnamefont {Atta-Fynn}}, \ and\
  \bibinfo {author} {\bibfnamefont {P.}~\bibnamefont {Biswas}},\ }\href
  {\doibase 10.1038/s41598-020-64327-3} {\bibfield  {journal} {\bibinfo
  {journal} {Sci. Rep.}\ }\textbf {\bibinfo {volume} {10}},\ \bibinfo {pages}
  {7742} (\bibinfo {year} {2020})}\BibitemShut {NoStop}%
\bibitem [{Note3()}]{Note3}%
  \BibitemOpen
  \bibinfo {note} {For recent, large-scale atomistic simulations using ML 
  potentials, see, e.g., Ref.\ \citenum{Smith2021}, or: W. Jia, H. Wang, 
  M. Chen, D. Lu, L. Lin, R. Car, W. E, and L. Zhang, in \textit{SC '20: 
  Proceedings of the International Conference for High Performance Computing, 
  Networking, Storage and Analysis}, 5 (2020).}\BibitemShut {Stop}%
\bibitem [{\citenamefont {Durandurdu}\ and\ \citenamefont
  {Drabold}(2001)}]{Durandurdu2001}%
  \BibitemOpen
  \bibfield  {author} {\bibinfo {author} {\bibfnamefont {M.}~\bibnamefont
  {Durandurdu}}\ and\ \bibinfo {author} {\bibfnamefont {D.~A.}\ \bibnamefont
  {Drabold}},\ }\href {\doibase 10.1103/PhysRevB.64.014101} {\bibfield
  {journal} {\bibinfo  {journal} {Phys. Rev. B}\ }\textbf {\bibinfo {volume}
  {64}},\ \bibinfo {pages} {014101} (\bibinfo {year} {2001})}\BibitemShut
  {NoStop}%
\bibitem [{\citenamefont {Glielmo}\ \emph {et~al.}(2018)\citenamefont
  {Glielmo}, \citenamefont {Zeni},\ and\ \citenamefont {{De
  Vita}}}]{Glielmo2018}%
  \BibitemOpen
  \bibfield  {author} {\bibinfo {author} {\bibfnamefont {A.}~\bibnamefont
  {Glielmo}}, \bibinfo {author} {\bibfnamefont {C.}~\bibnamefont {Zeni}}, \
  and\ \bibinfo {author} {\bibfnamefont {A.}~\bibnamefont {{De Vita}}},\ }\href
  {\doibase 10.1103/PhysRevB.97.184307} {\bibfield  {journal} {\bibinfo
  {journal} {Phys. Rev. B}\ }\textbf {\bibinfo {volume} {97}},\ \bibinfo
  {pages} {184307} (\bibinfo {year} {2018})}\BibitemShut {NoStop}%
\bibitem [{\citenamefont {Vandermause}\ \emph {et~al.}(2020)\citenamefont
  {Vandermause}, \citenamefont {Torrisi}, \citenamefont {Batzner},
  \citenamefont {Xie}, \citenamefont {Sun}, \citenamefont {Kolpak},\ and\
  \citenamefont {Kozinsky}}]{Vandermause2020}%
  \BibitemOpen
  \bibfield  {author} {\bibinfo {author} {\bibfnamefont {J.}~\bibnamefont
  {Vandermause}}, \bibinfo {author} {\bibfnamefont {S.~B.}\ \bibnamefont
  {Torrisi}}, \bibinfo {author} {\bibfnamefont {S.}~\bibnamefont {Batzner}},
  \bibinfo {author} {\bibfnamefont {Y.}~\bibnamefont {Xie}}, \bibinfo {author}
  {\bibfnamefont {L.}~\bibnamefont {Sun}}, \bibinfo {author} {\bibfnamefont
  {A.~M.}\ \bibnamefont {Kolpak}}, \ and\ \bibinfo {author} {\bibfnamefont
  {B.}~\bibnamefont {Kozinsky}},\ }\href {\doibase 10.1038/s41524-020-0283-z}
  {\bibfield  {journal} {\bibinfo  {journal} {npj Comput. Mater.}\ }\textbf
  {\bibinfo {volume} {6}},\ \bibinfo {pages} {20} (\bibinfo {year}
  {2020})}\BibitemShut {NoStop}%
\end{thebibliography}
\end{document}